\documentclass[a4paper,11pt]{article}
\pdfoutput=1 

\usepackage{jcappub} 
\usepackage{graphicx}
\usepackage{subcaption}                 
\usepackage[T1]{fontenc} 
\usepackage{amsmath}
\usepackage{multirow}
\usepackage{ulem}
\definecolor{mycolor}{rgb}{1.0, 0., 1.0}
\usepackage[dvipsnames]{xcolor}

\usepackage{xspace}
\newcommand{\rose}{Mff$\epsilon_{009}$-MecFB\xspace}
\definecolor{colorROSE}{HTML}{DE1562}

\newcommand{\rouge}{Mff$\epsilon_{100}$-MecFB\xspace}
\definecolor{colorRED}{HTML}{8B0000}

\newcommand{\blue}{Mff$\epsilon_{009}$-DCool\xspace}
\definecolor{colorBLUE}{HTML}{16A9D1}
\newcommand{\verte}{Mff$\epsilon_{100}$-DCool\xspace}
\definecolor{colorGREEN}{HTML}{006400}
\newcommand{\jaune}{KSlaw-DCool\xspace}
\definecolor{colorYELLOW}{HTML}{CCAA11}

\title{\boldmath Cosmological simulations of the same spiral galaxy: satellite properties, the role of baryonic physics and star formation history in shaping dark matter cores/cusps.}

\author[a,b]{A. Nu\~nez-Casti\~neyra,}
\author[c]{E. Nezri,}
\author[c,d]{P. Mollitor,}
\author[c]{L. Michel-Dansac,}
\author[e]{J. Devriendt,}
\author[f]{R. Teyssier.}

\affiliation[a]{Université Paris-Saclay, Université Paris Cité, CEA, CNRS, AIM, 91191, Gif-sur-Yvette, France }
\affiliation[b]{Institut d’Astrophysique de Paris, Sorbonne Université, CNRS, UMR 7095, 98 bis boulevard Arago, 75014 Paris, France}

\affiliation[c]{Aix Marseille Univ, CNRS, CNES, LAM, Marseille, France 
}

\affiliation[d]{Department of Physics and Materials Science, Faculty of Science, Technology and Medicine, University of Luxembourg, 162 A, Avenue de la Faïencerie, 1511 Luxembourg, Luxembourg}

\affiliation[e]{Sub-department of Astrophysics, University of Oxford, Keble Road, Oxford OX1 3RH, UK}

\affiliation[f]{Department of Astrophysical Sciences, Princeton University, Princeton, NJ 08540, USA}

\emailAdd{nunezcas@iap.fr}
\emailAdd{emmanuel.nezri@lam.fr}
\emailAdd{pol.mollitor@education.lu}
\emailAdd{leo.michel-dansac@lam.fr }
\emailAdd{julien.devriendt@physics.ox.ac.uk}
\emailAdd{romain.teyssier@gmail.com}

\abstract{We investigate the role of baryonic physics in shaping the population, structure, and internal dynamics of galactic subhalos using the Mochima suite of cosmological zoom-in simulations. A refined method is developed to identify bound subhalo material by isolating the local gravitational potential and applying multi-criteria phase-space selection. This approach enables a robust characterisation of subhalo properties across five baryonic runs with varying prescriptions for star formation, and supernova and protostellar feedback, as well as a dark matter-only baseline. At the population level, we find that host halo concentration, modulated by baryonic feedback, is a key predictor of subhalo survival. Subhalos with more massive stellar components exhibit deeper internal potentials and enhanced resilience to tidal disruption. At the structural level, we identify a broad diversity in inner dark matter profiles, consistent with observations of dwarf galaxies. We show that this diversity correlates with both star formation history and environmental interaction. In particular, galaxies that form most of their stars early tend to retain steep cusps, while those with extended or recent star formation exhibit oscillating inner slopes shaped by bursty feedback and tidal perturbations. These findings suggest that the so-called "diversity problem" may reflect the complex interplay between feedback history and gravitational environment, rather than a breakdown of cold dark matter predictions.
  }

\begin{document}
\today
\maketitle

\flushbottom
\section{Introduction}\label{sec:intro}

In a $\Lambda$CDM universe, galaxies form hierarchically within massive dark matter structures known as halos \cite{WhiteRees1978, WechslerTinker2018}, which merge over time to build larger systems. More massive halos are typically surrounded and populated by numerous smaller subhalos that evolve under the influence of the massive halo, their host. The evolutionary paths of subhalos and the satellite galaxy they host are often complex and dynamic. Initially forming as independent systems and later accreted by more massive galaxies these halos are drawn into various possible orbital trajectories, ranging from slow passages through the outskirts of the host halo to rapid and disruptive encounters near its centre. These orbits are non-Keplerian due to the effects of dynamical friction and tidal forces. As a satellite orbits its host, tidal forces strip the mass from the outskirts of its subhalo. This mass loss progressively reduces the region shielded from the tidal field of the host. At the same time, tidal shocks during pericentric passages heat the interior of the subhalo, pushing material outward toward its shrinking tidal boundary \cite{Hayashi2003}(see also \cite{Moore2000} for a historical overview).

The study of subhalos lies at the centre of many unresolved questions in cosmology. Their complex dynamical evolution poses a major systematic challenge for several advanced cosmological probes. Subhalos populations reflect galaxy populations and are connected to gravitational lensing signals, and dark matter annihilation predictions. These are non-linear systems that typically require simulations for accurate modelling. However, they are deceptively difficult both to simulate and to reliably identify in post-processed data. As a result, uncertainties in their survival and disruption limit the predictive power of theoretical models (see \cite{Onions2012, Mansfield2024} for a detailed description and verification on current methods). Some simulation work also highlights the mass and orbit-dependent competition between quenching and disruption, with ultra-faint satellites often quenching before infall while more massive systems undergo disruption prior to star formation shutdown \citep{Pathak2025}. This tension is particularly evident in the too big to fail problem, where dark-matter-only simulations predict overly dense subhalos that are inconsistent with the observed kinematics of bright dwarf galaxies, underscoring the need to incorporate baryonic physics. On one hand, their dominant presence in the central regions can reshape the overall subhalo population-including the most massive systems-through tidal and gravitational effects; on the other, sufficiently strong feedback episodes can reduce central dark matter densities within subhalos, bringing them closer to the masses inferred for observed satellites\cite{Boylan-Kolchin2011}.

The mass distribution within subhalos, as inferred from galaxy rotation curves, raises important questions about the existence of dark matter, its dynamics in the centres of dwarf galaxies, and, particularly, whether a universal inner density profile exists \cite{deBlok2008}. In fact, the matter of the inner profile of dark matter halos is far from settled. Simulations including cold dark matter tend to favour centraly growing density profile, or cusps, while high-resolution observations lead to the inferences of a mix between cusps and centrally constant density profiles, cores, with the latter being slighly more common \cite{Oh2011}. This is known as the core-cusp (CC) problem \cite{Moore1994,Flores1994} (see \cite{DelPopolo2021} for a review) or maybe more accurately the diversity problem \cite{Oman2015}. The formation of cusps and the challenges associated with satellite galaxies in $\Lambda$CDM are closely linked. Some unified models propose mechanisms that both transform cusps into cores and help alleviate other problems associated with satellites \cite{Zolotov2012,DelPopolo2014}, since the shape of the host halo significantly influences its tidal impact on satellites \cite[e.g.,][]{Mashchenko2006,Mashchenko2008,Penarrubia2010,Errani2023}, with cored profiles in the host galaxy even capable of completely disrupting satellite systems \cite{Penarrubia2010}. A similar mass-dependent structural trend is observed in TNG50 galaxies, where low-mass systems typically host dispersion-supported clumps while more massive galaxies develop extended, rotationally supported disks \citep{Celiz2025}.

There are various approaches to the CC problem and satellite populations; from considering alternatives to cold dark matter \cite{Sommer-Larsen2001,Peebles2000,Goodman2000,Hu2000,Cen2001,Kaplinghat2000,Spergel2000}, altered DM power spectrum \cite{Zentner2003}, modifications of newtonian dynamics (MOND) \cite{Milgrom1983a,Milgrom1983b} and of general relativity \cite{Buchdahl1970,Starobinsky1980,Bengochea2009,Dent2011}, all the way to revision of the shortcomings of simulations themselves. For example, overmerging due to unresolved relaxation \cite{deBlok2001,deBlok2003,Power2002} and the dynamics of baryons which are expected to be the dominant component for smaller galaxies. This work focus on uncertainties induced by the latter.  Several explanations have been proposed for the observed diversity in rotation curves and the presence of cores and/or cusps. Some studies suggest a correlation with stellar mass fraction \cite{DiCintio:2013qxa, Tollet2016,Chan2015}, while others attribute it to gas dynamics driven by stellar feedback \cite{Jackson2024, Read2016}. However, the explanation of the stellar mass fraction has been challenged by results from other cosmological simulations \cite{Bose2019}, which emphasize the impact of various sources of uncertainty, including the galaxy-to-galaxy variation, as further demonstrated by differences in the APOSTLE and Auriga simulations \cite{Richings2020}. The idea that the inner DM slope is somehow related to the baryonic content is in line with what is observed in the THING survey where low-surface brightness galaxies seem to favor cored profiles while luminous galaxies are equally well fitted by profiles with a cusp and profile with a core.
On the simulation side, uncertainties in the numerical modelling of sub-resolution physical processes complicate direct comparisons between simulations using different prescriptions. These modelling uncertainties must be carefully considered when interpreting simulation outcomes. The Mochima simulation\cite{Nunez-Castineyra:2020ufe} offers a valuable case study in this context, as it features the same galaxy re-simulated with different sub-resolution models for star formation and supernova feedback , thus contributing to the understanding of model-dependent numerical uncertainties.
The formation and evolution of satellite galaxies around Milky Way (MW)-like systems offer valuable insight into galaxy formation models and the nature of dark matter. Observations of satellite systems in the MW and Andromeda (M31) reveal complex distributions of stars and dark matter that theoretical models aim to reproduce \cite{Tollerud2008,McConnachie2012}. Numerical simulations serve as powerful tools for exploring these systems, enabling controlled studies of how different physical processes influence galaxy evolution \cite{Fattahi2016,Dubois2021NH,Grand2017}. Previous analyses of the Mochima simulations have focused on the baryonic distribution \cite{Nunez-Castineyra:2020ufe} and the dark matter content of the primary galaxy \cite{Nunez-Castineyra2023}, these works will be refered to as Paper 1 and Paper 2 from now on. However, the distribution and properties of satellite galaxies in these simulations remain unexplored. Satellite systems are expected to be particularly sensitive to both internal processes, such as supernova feedback, and external factors like tidal interactions with the central galaxy (e.g., \cite{Wetzel2016}). Understanding how variations in subgrid physics affect these interactions is crucial for interpreting observational data.

The structure of the paper is as follows. The simulation and post-processing methods are described in Section \ref{sec:simus}.  The features of the subhalo populations are explored in Section \ref{sec:GalacticSubhalos}. The discussion is provided in Section \ref{sec:Discussion}, and the summary and conclusions are presented in Section \ref{sec:summary}.




\section{Simulations and methods}\label{sec:simus}

The following analyses focus on the subhalos within the dark matter halos from the simulations presented in Paper 1, where the same halo is simulated six times: one run comprises only dark matter (DMO), and the other five include hydrodynamics (hydro) runs, all resulting in a spiral galaxy named the \emph{Mochima} galaxy. These simulations are conducted using the AMR code \textsc{RAMSES} \cite{Teyssier:2001cp}, starting from identical initial conditions generated with the MUSIC package \cite{HahnAbel2013}, within a cubic cosmological box of approximately 36 Mpc on each side. The zoom-in simulations employ a nested structure of five convex-hull volumes with varying dark matter resolutions (particle mass), ranging from a resolution equivalent to $128^3$ particles in the outermost region to $2048^3$ particles in the innermost volume. The innermost and most resolved layer, corresponding to the decontaminated\footnote{The decontamination process is performed using the public HAST package \url{https://bitbucket.org/vperret/hast/wiki/Home}} Lagrangian volume of the final galactic halo, contains dark matter particles with a mass of approximately $1.9\times10^5$ M$_{\odot}$.

While the five hydro runs share identical initial conditions, they incorporate different numerical models for star formation and supernova feedback, as detailed in Paper 1. This suite includes one control run using benchmark baryonic physics implementations from previous simulations \cite{Mollitor:2014ara,Marinacci:2013mha}, and four additional runs that combine recently introduced models for star formation (SF) and supernova (SN) feedback. The labels and main prescriptions are as follows. The control run, labeled KSlaw-DCool, employs star formation based on the Kennicutt-Schmidt law (KSlaw) \cite{Kennicut1998} and applies the Delayed Cooling (DCool) prescription for supernova feedback \cite{Teyssier2013}. This setup is compared with two alternative approaches for star formation. In the first modification, the star formation strategy transitions to the multi-freefall (Mff) model \cite{Federrath2012}, which is based on turbulent, magnetized molecular clouds \cite{Krumholz2005}. This prescription computes the star formation efficiency in each gas volume element considering its surrounding environment. Inside this calculation it introduces a multiplying variable, $\epsilon$, that adjusts the local star formation efficiency, for which two extreme cases have been chosen: a strong protostellar feedback scenario where $\epsilon=0.09$ and a weak protostellar feedback scenario where $\epsilon=1$. These runs are labeled Mff$\epsilon_{009}$-DCool and Mff$\epsilon_{100}$-DCool, respectively. In the final two runs, the SN feedback prescription is modified to implement a model based on the Sedov-Taylor phases of the supernova explosion, referred to as Mechanical feedback (MecFB) \cite{Kimm2015}. This results in the runs labeled Mff$\epsilon_{009}$-MecFB and Mff$\epsilon_{100}$-MecFB for the strong and weak protostellar feedback assumptions, respectively.

In all cases, the final galaxy is a disc galaxy with a substantial central bulge, although the relative mass of the bulge varies depending on the simulation. At redshift zero, the mass of the dark matter halo ranges between $0.92\times10^{12}$ M${\odot}$ and $1.13\times10^{12}$ M${\odot}$, depending on the run (see Table 1 of Paper 1). These halos exhibit a quiet merger history, with no major mergers for $z<2$, and are located within a filament near a massive neighbor of approximately $1.0\times10^{13}$ M$_{\odot}$, located about 6 Mpc away.

\begin{figure}[t]
\centering
\begin{subfigure}[b]{0.82\linewidth}
\includegraphics[width=\linewidth]{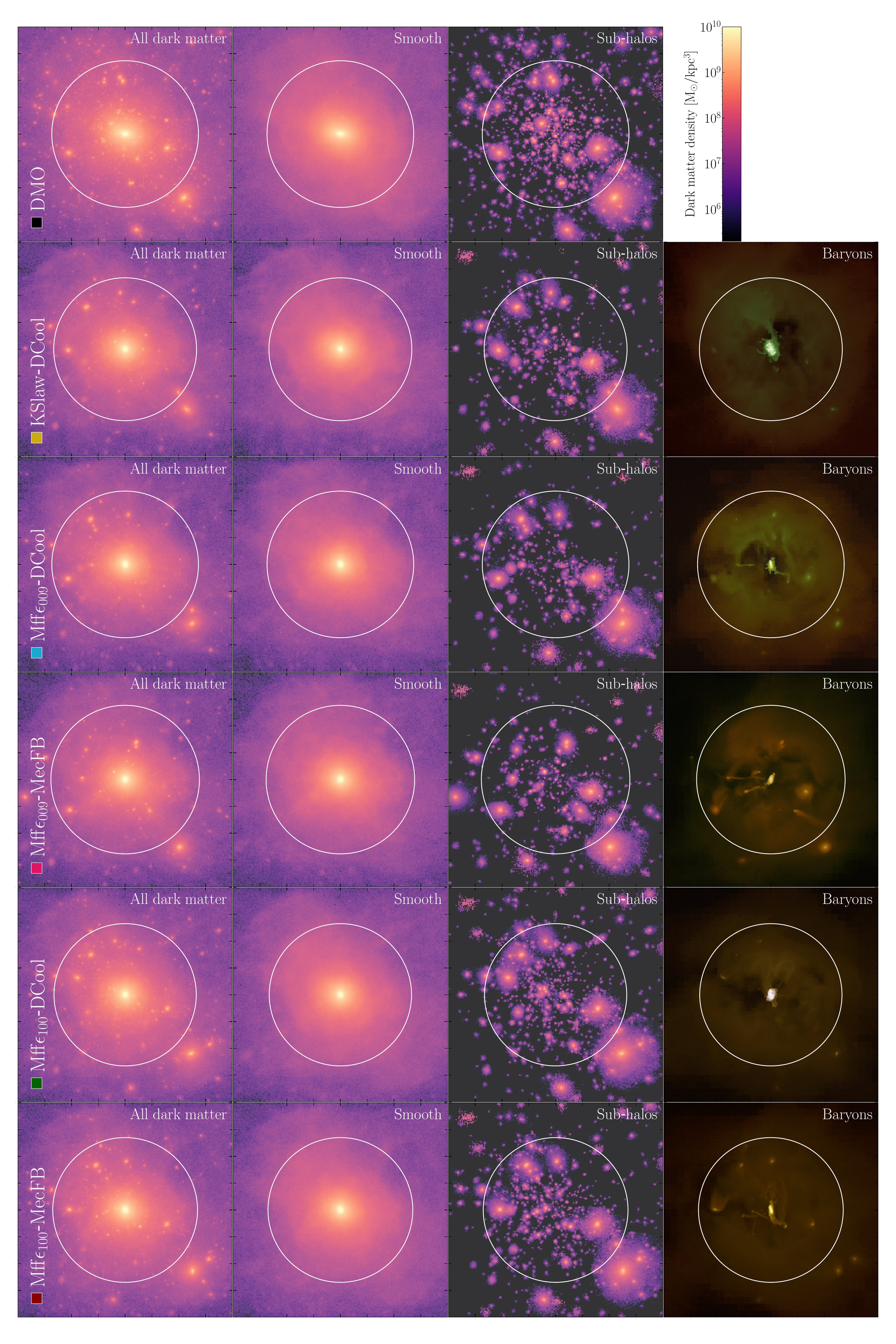}
\end{subfigure}

\caption{Decomposition of the dark matter and baryonic components in the six simulation runs. Each row shows, from left to right, the total dark matter density map, the separation into the smooth dark matter component and the subhalo-bound component, and, when available, the baryonic distribution obtained through radiative transfer imaging with SKIRT \cite{SKIRTCamps2015}.}\label{fig:subhalosmaps}
\label{fig:mapsfull}%
\end{figure}
\subsection{Subhalo detection}
Detecting subhalos in numerical simulations is a complex task that requires careful consideration of both the simulation outputs and the definition of a subhalo. A subhalo is a self-gravitating structure that remains under the gravitational influence of a more massive halo. A first-order approach to its identification is typically performed in configuration space by locating local overdensities, which serve as the centres of substructures. The subsequent challenge is to define the boundary of the substructure. If a subhalo were assumed to be spherical, which is rarely the case, a single radius could define this boundary. However, several definitions exist for the radius that marks the extent of the subhalo.

\paragraph{Density} One possible approach is to calculate the spherically averaged density profile of the subhalo and identify the radius at which the profile falls below the local background density of the host halo at the position of the subhalo (see section 4.1 and figure 14 of \cite{Springel:2008cc}. This ``truncation radius'' is intended to approximate the transition between the region where the subhalo dominates the gravitational field and the region where the host halo becomes dominant. However, since subhalos are not perfectly spherical and are subject to tidal forces from the host, the boundary is often distorted, with the side facing the host typically being more stripped and compressed. Consequently, the real transition is gradual and asymmetric, making this approach an approximation.

\paragraph{Mass} Another useful definition is the Jacobi radius or tidal radius, also known as the tidal radius, which represents the maximum distance from the subhalo centre within which mass remains gravitationally bound to the subhalo. In the case of a satellite (or clump) embedded in a smooth, spherically symmetric background at distance \( R \) from the centre the tidal radius is given by:
\begin{equation}\label{eq:rtidal}
r_{\rm tidal} \equiv R \left( 
\frac{m_{\mathrm{int}}(r_j)}{3 M(R) \left(1 - \frac{1}{3} \frac{\mathrm{d} \ln M}{\mathrm{d} \ln R}(R) \right)} 
\right)^{1/3}.
\end{equation}

Here, the function \( m_{\mathrm{int}}(r_j) \) denotes the internal mass of the satellite enclosed within \( r_j \), while \( M(R) \) is the mass of the host halo enclosed within radius \( R \). The term \( \frac{\mathrm{d} \ln M}{\mathrm{d} \ln R}(R) \) represents the logarithmic slope of the mass profile of the host halo, evaluated at \( R \). This definition is nonetheless an approximation, as it corresponds to the location along the line connecting the subhalo and the host halo where the gravitational influence of the subhalo balances that of the host. However, the surface that fulfills this condition is not spherical. Additionally, the calculation of the Jacobi radius typically assumes a static host potential and that the subhalo follows a circular orbit, conditions that may not be strictly satisfied in a cosmological environment. Moreover, calculating the Jacobi radius requires additional assumptions, such as a host halo with a spherically symmetric but non-homogeneous mass distribution, satisfying $M \propto r^{\alpha}$ with $\alpha < 3$ \cite{binney2011galactic}.

\paragraph{Gravitational potential} An alternative method uses the gravitational potential generated by all simulation components at the position of each dark matter particle. By finding the point along the line connecting the centres of the subhalo and the host halo where the gravitational potential reaches a maximum, one can define the edge of the subhalo as the distance from its centre to this point, denoted $r_{\rm max}$. This approach differs from the Jacobi (tidal) radius definition, as it relies purely on the gravitational potential without explicitly accounting for centrifugal forces or assuming a circular orbit. It provides a more direct, numerical characterisation of the transition between the subhalo and the host halo, based on the actual mass distribution at the time of analysis, without requiring spherical symmetry or analytic approximations.

However, these approaches remain approximate, as neither the subhalos nor the host halos are perfectly spherical, particularly in the outer regions of the host halo (see Paper 2). Another limitation is that these methods generally assume that all particles within a given radius are bound to the subhalo, thereby ignoring particles belonging to the main halo that are erroneously enclosed.

To overcome the limitations imposed by spherical symmetry, several grouping algorithms have been developed that operate purely in configuration space or in phase space, starting from density peaks. Examples include the Friends-of-Friends (FoF) algorithm \cite{HuchraGeller1982,Davis1985} and the HOP algorithm (see \cite{Maciejewski2009,Behroozi2012Rockstar}), which rely on proximity criteria. However, these methods still require the calibration of free parameters that lack a clear physical justification.
An alternative approach relies on the full phase-space information by evaluating the potential energy, $U$, and kinetic energy, $K$, of each particle in the reference frame of the subhalo. Some methods go as far as to use machine learning to distinguish orbiting from infalling dark matter particles based on their phase-space properties \citep{Vladimir2025}. In the first approach, the kinetic energy can be computed relative to the centre-of-mass velocity of the subhalo to accurately assess the internal dynamical state. Then, the particles are then considered bound if they have a negative total energy, i.e., if $|U| > K$. This criterion is effective when the gravitational potential is computed using only the local environment around the subhalo, rather than the full host-subhalo system. Including the gravitational contribution from the host halo can otherwise lead to the misidentification of genuinely bound particles as unbound. This procedure forms the core of the SUBFIND algorithm \cite{Springel2001SUBFIND}, which has been widely used to identify gravitationally self-bound structures in cosmological simulations. However, this method can face challenges in crowded environments, where overlapping structures and tidal debris complicate the definition of individual subhalos \cite{Mansfield2024}. In such complex regions, tracking the evolution of particles across snapshots becomes essential to robustly follow subhalos approaching complete disruption. Nevertheless, single-snapshot population studies, such as the one presented here, remain valuable for characterizing the distribution of still-bound structures.


In what follows, building upon the discussions above, an extended approach was adopted to identify the material truly bound to subhalos. First, subhalos detected with the Rockstar halo finder \cite{Behroozi2012Rockstar} in the last snapshot of each simulation were used as the basis for this analysis. The substructures included were selected according to stringent conditions: their distance from the centre of the main halo was required to lie between 1 kpc and $R_{\rm max}$, the radius where the gravitational influence of the host halo ends (see Paper 2), and their mass had to exceed eighty\footnote{Similar studies establish a minimum number of particles to define a detectable subhalo on the order of tens of particles. For example, \cite{Onions2012} uses twenty particles, and \cite{Mansfield2024} adopts a threshold of around forty. In this study, a more conservative value was chosen.} times the mass of a dark matter particle, ensuring robust resolution.

The particles gravitationally associated with each clump are identified using a two-stage procedure. Strongly bound particles are selected based on an energy criterion combined with a directional constraint, while loosely bound particles in the outskirts are identified using a phase-space proximity measure relative to the core. Full details of these selection criteria are provided in Appendix \ref{app:particleselection}.

This multi-dimensional phase-space analysis naturally accounts for the asymmetry and tidal distortion experienced by subhalos, avoiding the pitfalls of assuming spherical symmetry or static host potentials. The resulting detections are illustrated in Figure~\ref{fig:subhalosmaps}, where, for each of the six simulation runs, the leftmost column shows the total dark matter density map, the middle columns show the decomposition into the smooth dark matter component (particles not identified as bound to any subhalo) and the subhalo component (particles identified as bound), and, when available, the rightmost column shows the baryonic distribution obtained through radiative transfer images generated with SKIRT \cite{SKIRTCamps2015}.

Overall, this method provides a significant improvement in the physical characterisation of subhalo material and in the calculation of their properties, with a precision that considerably exceeds that of simpler approaches. Although the procedure is more involved, the resulting clarity in the definition of bound substructure more than justifies the additional complexity. It must be noted, however, that the method is specifically designed to refine the characterisation of compact, gravitationally bound subhalos based on an initial detection performed by the Rockstar halo finder, and is not intended as a substitute for halo-finding algorithms. As a result, the method may not capture more diffuse or extended tidal features, such as stellar or dark matter streams, or faint remnants of recently disrupted substructures, which do not form a self-bound structure but can nevertheless retain dynamical coherence over large distances. A natural extension of this method would be to apply it across multiple simulation snapshots, tracking particles bound to the same structures over time, even after their disruption. While Section \ref{subsec:subhalohistory} presents an analysis of the time evolution of the subhalo population, the present work focuses exclusively on subhalos identified as gravitationally bound at redshift zero, and does not address the detection, characterisation, or evolution of streams or faint remnants. These studies will be pursued in future work.

\subsection{Baryonics physics in the Mochima simulations}\label{subsec:catergories}
The aim of the Mochima simulations is to investigate how variations in subgrid physics affect the evolution of a galaxy formed in a cosmological context. The first stage of this simulation suite consists of six runs: one dark matter only (DMO) simulation, and five hydrodynamical reruns that include baryons (gas and stars) and implement different combinations of subgrid models. The global properties of the central galaxy in each run are discussed in detail in Paper 1 (baryonic distributions) and Paper 2 (DM structure). Although the subgrid models act jointly and in a highly non-linear fashion to shape the evolution of galactic structures, three categorizations emerge as particularly informative for interpreting the properties of the satellite population:

\begin{itemize}
    \item \textbf{Supernova Feedback Models:} Two distinct models of SN feedback are implemented, as described in Paper 1. The first, Delayed Cooling, has a smoother local impact but is highly efficient at expelling gas from the galactic disc. This model is used in three runs: {\jaune}, {\blue}, and {\verte}. The second model, Mechanical Feedback, is more effective at displacing local gas but less efficient at heating the ISM, thereby allowing for increased star formation. It is employed in the \rose  and \rouge runs.

    \item \textbf{Protostellar Feedback Efficiency:} While \jaune lacks the turbulent star formation prescription, the remaining four runs include it. In this framework, the parameter $\epsilon$ acts as an external suppression factor, applied multiplicatively in the turbulence-based star formation efficiency $\epsilon_{\rm ff}$ (see Equation 8 in Paper 1), thereby reducing the predicted efficiency. The strong-feedback cases adopt $\epsilon = 0.09$, representing a substantial reduction. This value is not necessarily unphysical, as feedback from stellar winds and radiation is expected to expel a significant fraction of the surrounding gas in young star-forming regions \cite{Krumholz2005,Federrath2012}. The runs implementing this strong protostellar feedback are \blue and \rose. In contrast, \verte and \rouge correspond to the respective runs without protostellar feedback, where $\epsilon = 1$.

    \item \textbf{Stellar Mass Extreme Cases:} The survival of subhalos depends on the competition between their internal potential wells and the gravitational influence of the central galaxy. This introduces a third useful classification based on global star formation efficiency. In particular, the least efficient run, \jaune, forms a relatively light stellar disc with a shallow potential, while the most efficient, \rose, forms stars efficiently even within subhalos, allowing them to survive closer to the galactic centre.
\end{itemize}

\section{The features of the galactic subhalos}\label{sec:GalacticSubhalos}

The evolving distribution of subhalos arises from the interplay between several physical processes, spanning orbital dynamics on galactic scales down to the physics of stellar populations. As subhalos orbit within the central galaxy, the host gravitational potential heats their internal velocity distribution and strips material from their outskirts. The likelihood of survival against this gravitational harassment increases when the internal potential well of the subhalo is sufficiently deep. Coincidentally, the central regions of galaxies are gravitationally dominated by the presence and dynamics of baryons, underscoring the central role that subgrid physics plays in shaping the evolution and disruption history of orbiting galaxies within massive systems. While the depth of the gravitational potential is a relevant indicator of the ability of the host to disrupt subhalos, the shape of the potential—closely related to the concentration of the dark matter halo—also plays a critical role. A more concentrated mass distribution leads to stronger tidal forces over a wider range of radii, increasing the efficiency of stripping and disruption. The survival of subhalos is therefore influenced not only by the minimum of the potential but also by its radial structure.

\begin{figure}[t]
\centering
\begin{subfigure}[b]{0.45\linewidth}
\includegraphics[width=\linewidth]{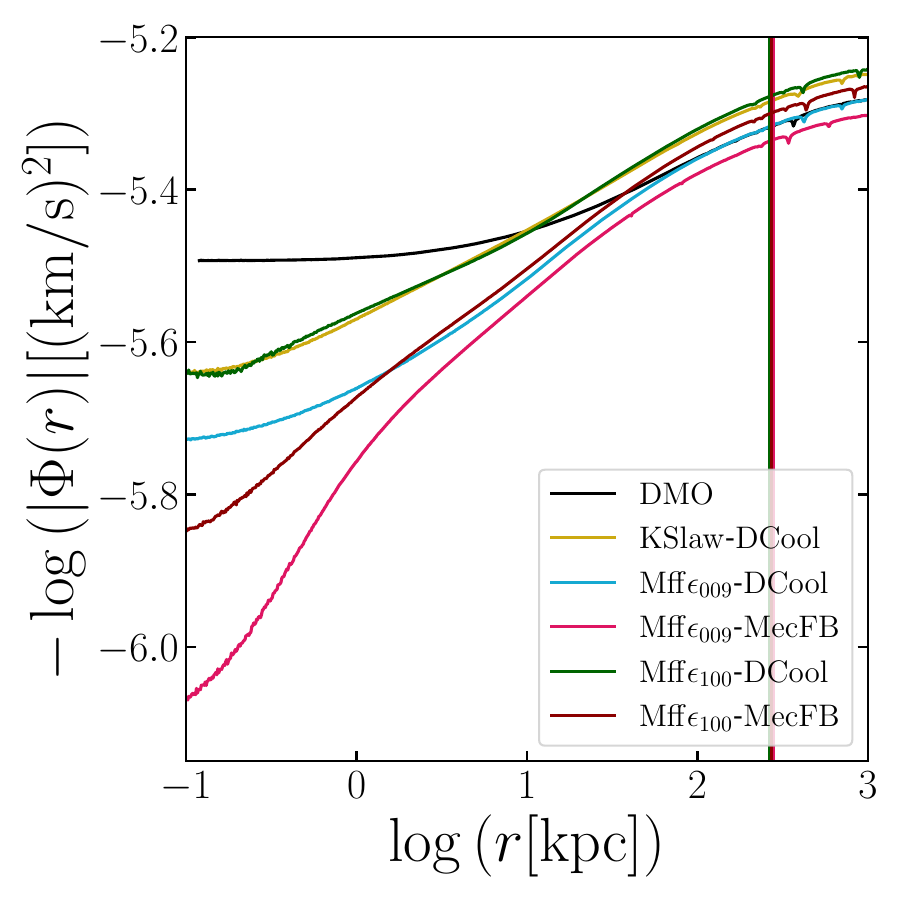}
\end{subfigure}
\begin{subfigure}[b]{0.45\linewidth}
\includegraphics[width=\linewidth]{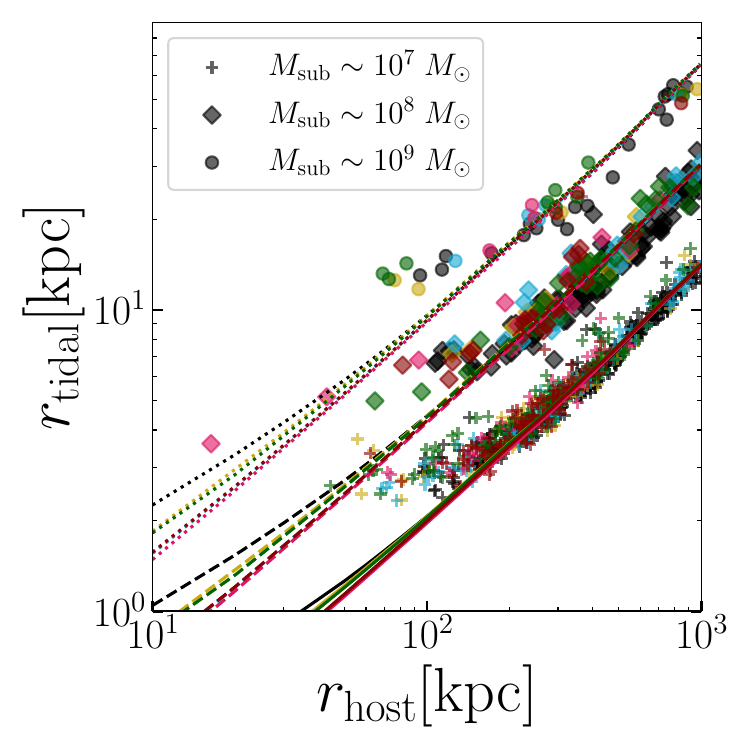}
\end{subfigure}
\caption{Left: Radial profiles of the gravitational potential for each run. Vertical lines indicate the virial radius of the main halo. Right: Tidal radius estimates for a subhalo on a circular orbit, computed for three subhalo masses ($10^7$~M$_\odot$ with crosses and solid lines, $10^8$~M$_\odot$ with diamonds and dashed lines, and $10^9$~M$_\odot$ with circles and dotted lines) using the enclosed host mass at each radius, $r_{\rm host}$. The markers correspond to the reconstructed mass of the detected subhalos and lines correspond to the analytical evolution of the tidal radius. The colors are consistent with each simulation color as in the left figure}
\label{fig:halofeatures}
\end{figure}

The left panel of Figure~\ref{fig:halofeatures} shows the radial profiles of the gravitational potential for each run, computed from all massive components: gas, stars, and dark matter. These profiles highlight two distinct regimes. In the central region, the presence of a baryonic disc deepens the potential well, with the hierarchy reflecting the supernova feedback implementation. The two runs with Mechanical Feedback (\rose and \rouge) exhibit the deepest central potentials, followed by the runs with Delayed Cooling. Within each feedback category, simulations with protostellar feedback produce deeper potentials than their counterparts without it.

In the outer halo region, where subhalos typically reside ($r \gtrsim 50$~kpc), the shallowest gravitational potentials are found in the two runs without protostellar feedback, as well as in the run with fixed star formation efficiency (\verte, \rouge, and \jaune). This trend helps explain why, in Figure~\ref{fig:mapsfull}, these three simulations—together with the DMO run—retain the largest number of surviving low-mass subhalos. The same argument can be framed in terms of a related structural feature: the concentration of the host dark matter halo. As shown in Table 1 of Paper 2, the simulations with the shallowest outer potentials are also those with the lowest concentrations\footnote{The concentration is defined as $c \equiv r_{200}/r_{-2}$, where $r_{200}$ is the radius enclosing a mean density 200 times the critical density, and $r_{-2}$ is the radius where the logarithmic slope of the density profile equals $-2$.}
, with values of $c = 16.3$, $20.7$, and $20.4$ for \verte, \rouge, and \jaune, respectively. The alignment between halo concentration and the depth of the outer potential provides a consistent framework for understanding subhalo survival in these runs, in agreement with results from  DMO large scale cosmological simulations \cite{Gao2011}.

The right panel of Figure~\ref{fig:halofeatures} illustrates the corresponding tidal radius estimates, $r_{\rm tidal}$ (equation \ref{eq:rtidal}), for subhalos of different masses. These estimates are based on a simple analytic model assuming a circular orbit and are computed from the enclosed host mass at a given radius using equation \ref{eq:rtidal} and shown with lines. They are compared with the tidal radii derived directly from the gravitational potential of each subhalo, shown on the scatter plot only for subhalos that are 10$\%$ close to the quoted mass. This radius is defined as the distance from its centre to the local maximum in the potential along the axis connecting it to the central halo. The two estimates diverge inside $r_{\rm host} \sim 100$~kpc, where the analytic tidal radius (lines) becomes extremely small, suggesting that either of the subhalos in this region are undergoing significant tidal disruption from the central potential and are therefore not found by our detection method, or that the circular orbit approximation is no longer valid.

\subsection{Stellar content of subhalos}
\begin{figure}[t]
\centering
\begin{subfigure}[b]{0.49\linewidth}
\includegraphics[width=\linewidth]{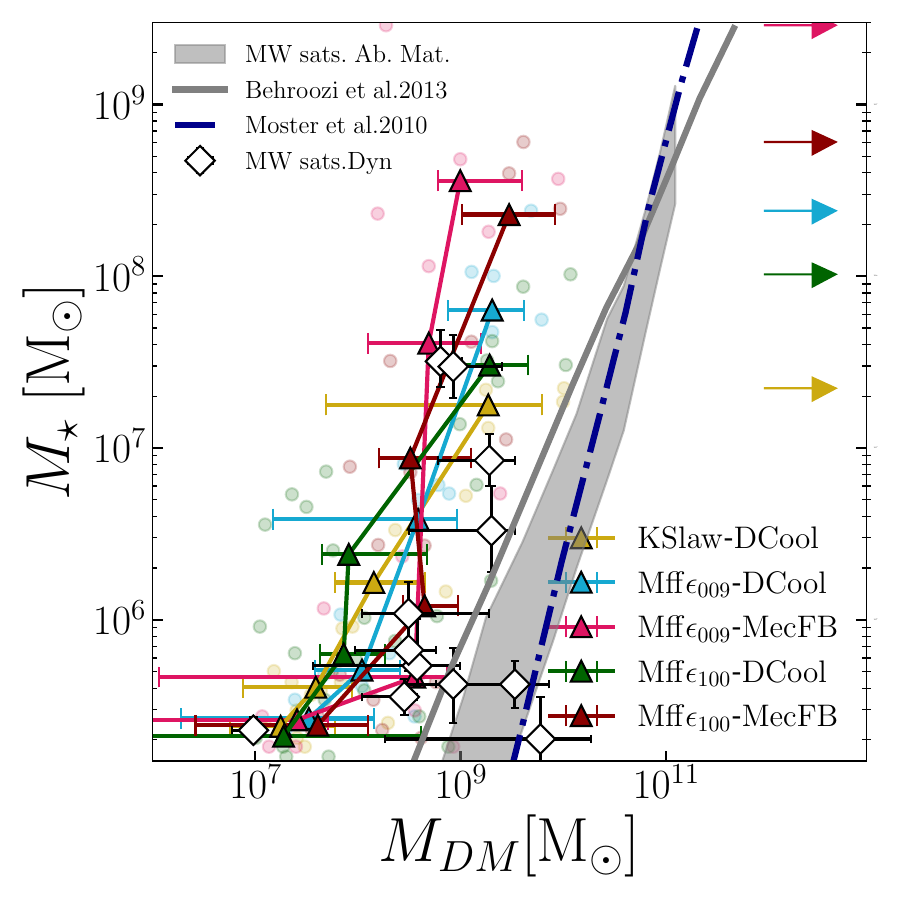}
\end{subfigure}
\begin{subfigure}[b]{0.49\linewidth}
\includegraphics[width=\linewidth]{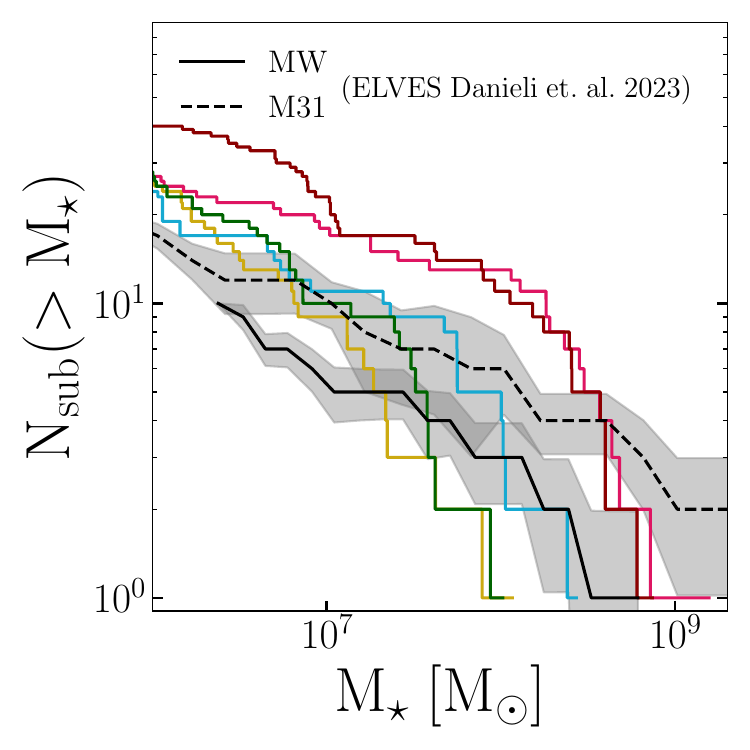}
\end{subfigure}
\caption{Left: Stellar-to-halo mass relation (SHMR) for subhalos in the Mochima simulations (circle markers). Triangles indicate the median and the 16th–84th percentile range in stellar mass bins. Black and white diamonds show dynamical inferences from \cite{Errani2018}, while the grey band represents abundance matching predictions for Milky Way satellites from \cite{Nadler2020}. Lines are extrapolations from central galaxy models by \cite{Behroozi2013} and \cite{Moster2010}. Arrows indicate the most massive stellar subhalo identified in each run. Right: Cumulative stellar mass distribution of satellites in the Milky Way and M31 from \cite{Danieli2023} (black lines), compared with subhalos in the five baryonic runs.}
\label{fig:stellarComps}
\end{figure}

Figure \ref{fig:stellarComps} shows the stellar content of subhalos in the five baryonic simulations. The left panel presents the stellar-to-halo mass relation (SHMR) of individual subhalos, compared against dynamical mass estimates for Milky Way satellites from \cite{Errani2018} and abundance matching constraints from \cite{Nadler2020}. Extrapolations of empirical SHMR relations for central galaxies are also shown, based on \cite{Behroozi2013} and \cite{Moster2010}. Coloured arrows indicate the most massive stellar subhalo in each run. Overall, the subhalos in the simulations exhibit stellar masses that are systematically higher than expected from observational constraints and abundance matching models, indicating that star formation may be overly efficient in these systems. In contrast, the total halo masses of the simulated subhalos show better agreement with the dynamical estimates for Milky Way satellites from \cite{Errani2018}\footnote{The estimates shown in Figure \ref{fig:stellarComps} correspond to those assuming central cusps. Adopting a cored profile would shift some of the inferred halo masses toward higher values, an effect that is less relevant for the more massive satellites, which are those well resolved in the simulations. Lower-mass ultra-faint dwarf galaxies are excluded from this comparison, as their stellar masses fall below the resolution limit (stellar particle mass $\sim 10^5$~M$_\odot$).}. In the cited work, the Milky Way's classical dwarf spheroidal satellites typically reside in halos with masses between $10^8$ and $10^{10}$~M$_\odot$, a range well covered by the resolved subhalo population in the Mochima simulations.
As subhalos orbit within their host, they lose mass—primarily dark matter—from their outer regions due to tidal stripping. This effect tends to shift their position leftward in the SHMR plane relative to central galaxies \cite{RodriguezPuebla2012}. In the simulations, this trend is particularly evident in the two runs with Mechanical Feedback, which also produce the most massive central stellar components. These runs yield subhalos with the highest stellar-to-halo mass ratios among all cases. This outcome likely reflects a combination of enhanced star formation efficiency within subhalos and the increased stripping efficiency caused by the deeper central potential of the host galaxy.

The right panel of Figure \ref{fig:stellarComps} presents the cumulative stellar mass distribution of subhalos. Black lines indicate the same metric for the Milky Way and M31 as reported in the Exploration of Local VolumE Satellites survey \cite{Danieli2023}. The scatter between the spectra of both galaxies is expected to reflect differences in formation times, as suggested by \cite{Applebaum2021}; in the present runs, this effect is isolated by construction, as all simulations share the same formation history, and the observed differences instead reflect the extent to which baryonic physics affects the subhalo populations. Reasonable agreement is observed between the simulations and the reference galaxies: \rouge and \rose display an excess of subhalos by less than a factor of two across most of the resolved mass range, along with a slope of the cumulative mass function that is similar to that inferred from observations. In particular, at higher stellar masses, these two runs (corresponding to those with Mechanical Feedback) succeed in populating subhalos with sufficient stellar mass to allow comparison with both reference galaxies. In contrast, runs with Delayed Cooling exhibit a deficit at stellar masses above $10^8$~M$\odot$, although comparable populations are produced below that threshold. However, the slope of the spectrum in these cases appears steeper than that inferred for the Milky Way and M31. Furthermore, the weak protostellar feedback cases (\verte and \rouge) are found to be more efficient at forming stars in low-mass subhalos ($M < 10^8$~M$\odot$) than their strong-feedback counterparts. These results support the validity of the simulated subhalo population in terms of observable stellar components. The following sections will examine aspects of the population that are either inaccessible to current observations or require more complex methodologies, beginning with the subhalo mass spectrum.

\begin{figure}[t]
\centering
\begin{subfigure}[b]{0.45\linewidth}
\includegraphics[width=\linewidth]{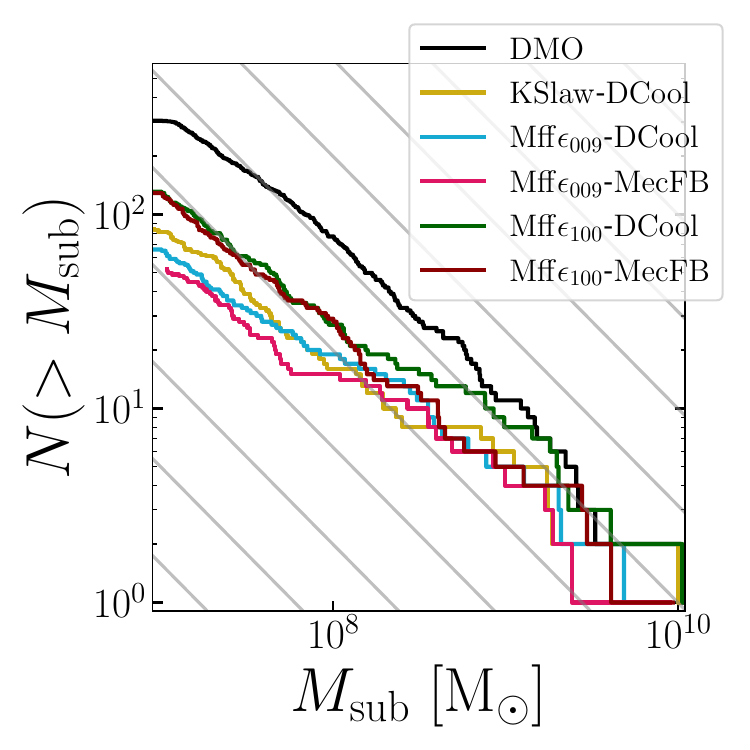}
\end{subfigure}
\begin{subfigure}[b]{0.45\linewidth}
\includegraphics[width=\linewidth]{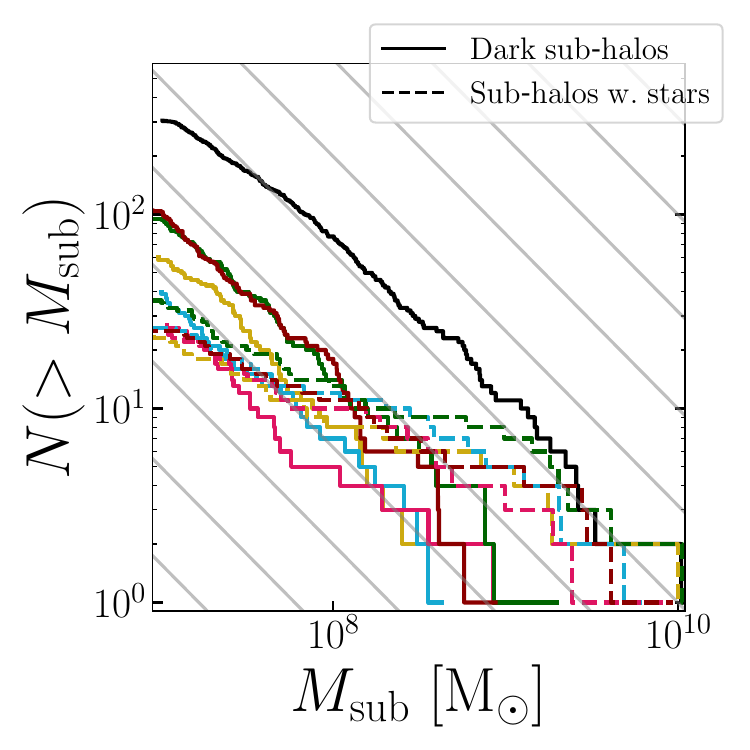}
\end{subfigure}
\caption{Cumulative subhalo DM mass spectra for the five runs, for all subhalos  left and  for separated population of subhalos without stars (solid lines)  and subhalos with stars (dashed lines) on the right. The gray lines show the slope consistent with $\frac{dN}{dM}\propto M^{-1.9}$ as observed in previous dark matter only simulations.}
\label{fig:Ncumulative}%
\end{figure}
\subsection{Mass spectrum}

Using the DMO run as a baseline, it is evident that the inclusion of baryons reshapes satellite demographics. The presence of baryons exerts a twofold impact on subhalo survival by increasing the efficiency of mass stripping through interactions with the central regions of the host mass distribution and by enhancing the internal gravitational binding of subhalos that contain a significant stellar component. On one hand, central baryons increase the ability of the host galaxy to strip mass from orbiting subhalos by creating a deeper and more concentrated central gravitational potential, which induces stronger tidal forces. In addition to the depth of this central potential (see left panel of Figure \ref{fig:halofeatures}), subhalos that pass through or near the baryonic disc may undergo disc shocking, wherein rapid changes in the vertical gravitational field inject energy into their constituent particle orbits. This process can further erode structural integrity and accelerate disruption, particularly for subhalos on plunging orbits through central regions. On the other hand, baryons residing within subhalos deepen their internal potential wells, thereby improving resistance to disruption, as a larger amount of energy is required to unbind their material.

Figure~\ref{fig:Ncumulative} presents the cumulative subhalo mass function. The left panel shows the full subhalo population, while the right panel separates subhalos based on their stellar content. The grey lines follow a slope consistent with $\frac{dN}{dM}\propto M^{-1.9}$ or $N(>M)\propto M^{-0.9}$, as expected from dark matter-only simulations \cite{Springel:2008cc,Diemand:2007qr,Wang2020}. This scaling reflects the near scale-invariance of hierarchical structure formation predicted by the Press–Schechter formalism \cite{PressSchechter1974} and its extensions \cite{Bond1991,LaceyCole1993}.

Although idealized models predict a slope close to $-2$, nonlinear processes such as tidal stripping and subhalo disruption tend to flatten this slope slightly, to values near $-1.9$ \cite{Hayashi2003,Tormen1998,Richings2020}. The concentration of the host halo is a key factor in this evolution, with more concentrated systems exerting stronger tidal fields that accelerate mass loss. Semi-analytical models incorporating these effects \cite{Zentner2007,SomervillePrimack1999} have demonstrated that environmental interactions are essential to shaping the subhalo mass function.

Differences in the subhalo mass function between runs-particularly those seen between \verte and \jaune-may arise from a combination of factors, including the properties of the central disc and the internal baryonic physics within subhalos. Variations in star formation and feedback models can affect the survival of clumps, especially in the mass range above $\sim 10^8$~M$_\odot$, where subhalos in the \verte run host a dominant stellar population. For lower-mass subhalos, disc shocking may contribute, although a quantitative assessment would require a dedicated orbital analysis and is beyond the scope of the present work.

The cumulative subhalo mass functions shown in Figure~\ref{fig:Ncumulative} reflect the influence of the depth of the central gravitational potential. The impact of this effect is evident in the runs with Mechanical Feedback, \rose and \rouge, where exceptionally deep central potentials are developed (see Figure~\ref{fig:halofeatures}). This results in efficient disruption of the most massive subhalos and, in the case of \rose—featuring the deepest potential due to an overly massive central stellar bulge—widespread disruption across the entire resolved mass range. A comparison between \verte and \jaune highlights a different regime, where the outcome is shaped by the balance between the depth of the central potential and the internal binding energy of subhalos due to their stellar content. Both galaxies possess very similar baryonic masses (see Figure 2 of Paper I) and comparable central potentials (see Figure~\ref{fig:halofeatures}), yet \verte shows limited disruption, particularly of massive subhalos ($M > 5 \times 10^8$~M$_\odot$), while \jaune shows significantly stronger depletion in this mass range. This contrast may arise because massive subhalos in \verte are more tightly bound internally, hosting more stellar mass and thus resisting tidal disruption more effectively. In contrast, subhalos in \jaune, though subjected to a similar host potential, are more weakly bound internally and are more readily disrupted. The subhalo population in \blue is likely shaped by a combination of both mechanisms: the central potential lies between those of the other simulation groups, and the disc is very massive relative to its central bulge, resulting in efficient disruption across all subhalo masses.

Additionally, the survival of low-mass subhalos appears to be correlated with the concentration of the host dark matter halo (see Table~1 of Paper II). Systems with higher concentrations exhibit denser central regions, which amplify the gravitational tides experienced by infalling subhalos. As a result, low-mass subhalos entering these environments are more efficiently disrupted, particularly during pericentric passages. This trend suggests that, beyond the direct influence of baryons, the structural properties of the host halo also contribute to shaping the subhalo population. In particular, higher concentrations steepen the gravitational potential gradient encountered by subhalos, enhancing the efficiency of tidal stripping and reducing the survival rate of the smallest systems. This scenario, when combined with the mechanisms described above, highlights the interplay between baryonic and dark matter-driven processes and emphasizes the complexity of subhalo evolution in cosmological environments.

The right panel of Figure~\ref{fig:Ncumulative} distinguishes subhalos that contain at least 0.1$\%$ of their mass in stars (dashed lines) from those without stellar content (solid lines). At high subhalo masses, stellar content is nearly always present, while low-mass subhalos are typically dark. This is not unseen and can be explained by the suppression of gas accretion in shallow potential wells following reionization, as shown in early simulations \citep{Chan2015}. Recent high-resolution results from the NewHorizon simulation confirm that most low-mass subhalos never form stars due to their inability to self-shield against the ionizing background \citep{Jeon2025}. In the case of the Mochima simulations, the majority of subhalos removed by tidal disruption in the concentrated runs belong to the dark, low-mass population. In contrast, massive subhalos with non-negligible baryonic content are more likely to survive, and may undergo structural modifications due to tidal interactions or internal feedback. Additionally, the slope in the cumulative mass function does not reproduce the behaviour expected by dark matter only simulations. The high-mass regime is where baryonic processes decisively influence subhalo survival. Although the reduction in the high-mass subhalo population is not significant in the \verte run, it is still clearly present -albeit less pronounced than at lower masses- in the other runs, indicating that the ``too big to fail'' tension is alleviated to varying degrees. These massive systems are also the best resolved in the simulations, which is why the analysis that follows focuses on this subset of systems, the regime of dwarf galaxies orbiting a MW-like galaxy.

\subsection{The history of the inner baryonic content and the dark matter halo of satellite galaxies}\label{subsec:subhalohistory}

The internal dynamics of dwarf galaxies pose a longstanding challenge to cold dark matter models, which generically predict cusps, while observations reveal a diversity of inner slopes in their dark matter halos. Bursty star formation and supernova-driven gas outflows have been proposed as mechanisms capable of flattening cusps into cores \cite{Governato:2009bg,Pontzen2012,El-Zan2016}, with recent work highlighting the duration of star formation as a key factor regulating this process \cite{Celiz2025}. Further extensions of these ideas have been explored by analytic models of burst-driven core formation at intermediate halo masses ($M_{\rm vir} \sim 10^{10}$–$10^{11}\,M_\odot$), claiming that stellar feedback in this range is energetic enough to modify the dark matter potential \citep{Freundlich2020,Li2023}.

Simulations of field dwarf galaxies have suggested a correlation between the inner dark matter slope and the stellar-to-halo mass fraction, with cores appearing preferentially in systems where stellar masses account for between 0.1$\%$ and 2$\%$ of the total halo mass \cite{DiCintio:2013qxa,Tollet2016,Tollet2017,DiCintio2017}. However, this relation has not been robustly confirmed across different simulation datasets \cite{Bose2019,Jackson2024}. Independent studies based on both cosmological simulations \cite{Fitts2017,Dekel2021} and idealized isolated galaxies \cite{Read2016,Hashim2023} support the idea that dark matter responds dynamically to repeated gas outflows. These studies demonstrate that bursty star formation can expel central gas and alter the orbits of dark matter particles, leading to core formation. While the precise thresholds vary, such effects are typically observed in halos with total masses between  $10^{10.5}<$~M/M$_{\odot}<10^{11.5}$.

The present analysis focuses on halos within the zoom-in region of the Mochima simulations, with total masses between $10^8$ and $5\times10^{10}$~M$_{\odot}$ and located within the gravitational region of influence of the main galaxy, defined as $R_{\rm max}$. While these are not field galaxies, many of the selected systems undergo minimal interaction with the central galaxy, particularly those observed on their first infall. This selection yields approximately 8–9 resolved subhalos per baryonic simulation. Although their masses fall below the typical mass range ($M_{\rm vir} \sim 10^{10}$–$10^{11}\,M_\odot$) considered in models of feedback-driven density profile transformation \citep{Pontzen2012,Freundlich2020,Li2023}, it is likely that similar effects are present in lower-mass subhalos that host a non-negligible stellar population.

The distinction between cusps and cores likely oversimplifies the complexity of inner halo structure. As shown by \cite{Oman2015}, dwarf galaxies exhibit a wide diversity of central dark matter slopes, indicating that baryonic processes—such as star formation history, assembly timing, and environmental interactions—can generate a spectrum of inner density profiles rather than a simple bimodal classification.

\begin{figure}[t]
\centering
\begin{subfigure}[b]{0.95\linewidth}

\includegraphics[width=\linewidth]{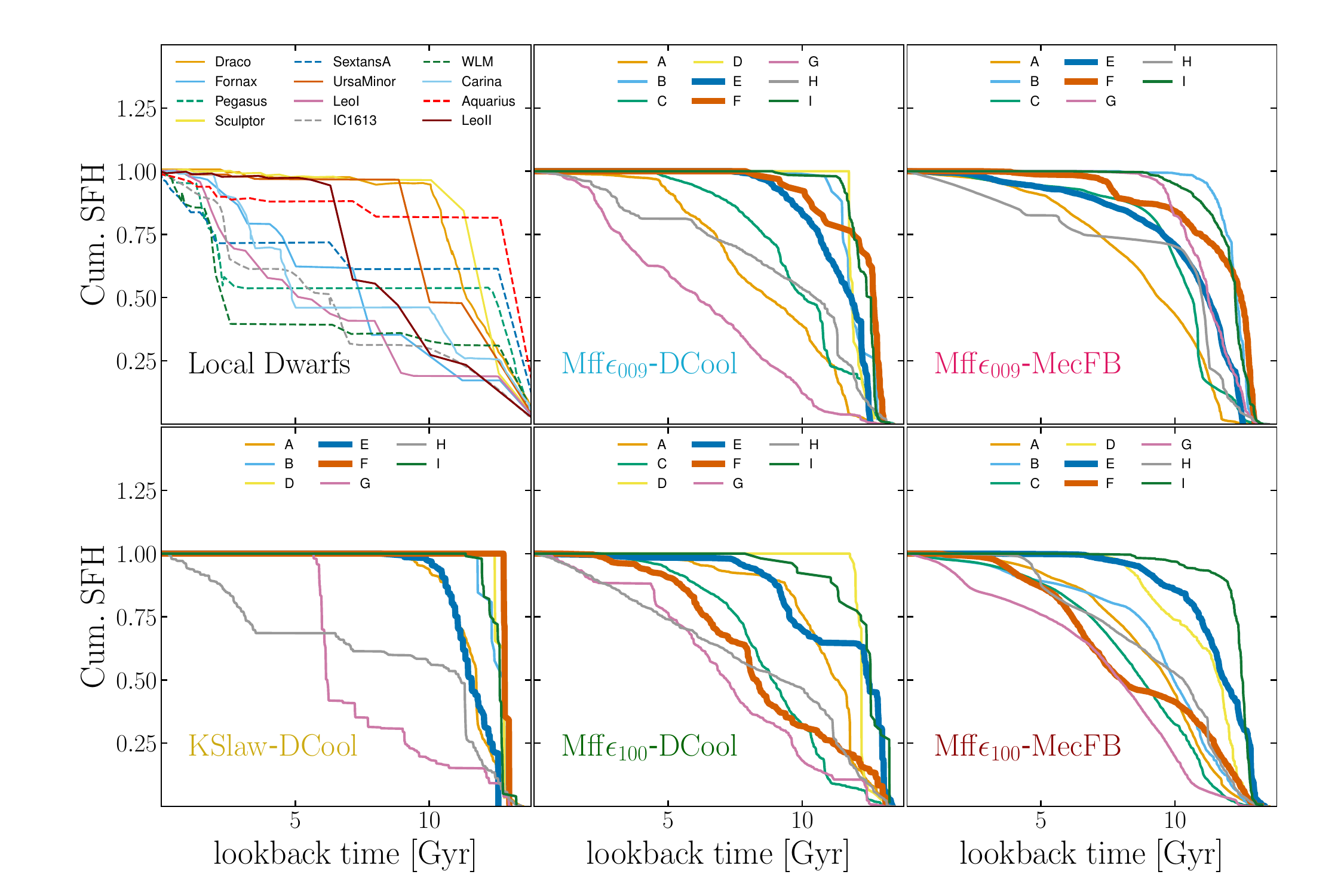}
\end{subfigure}

\caption{Cumulative star formation histories for dwarf galaxies. The upper-left panel shows observational data for Local Group systems from \cite{Weisz2014}, where solid lines represent Milky Way satellites and dashed lines denote more isolated dwarfs. The remaining panels show subhalos from the five baryonic simulations of the Mochima galaxy, selected within the region of influence of the central halo. The two thick lines are meant to highlight the two satellites detailed in Figure \ref{fig:evo}. The varying number of curves per panel reflects differences in subhalo disruption across runs.}
\label{fig:SFHall}%
\end{figure}

In this context, the star formation history (SFH) emerges as a valuable diagnostic. While gas content fluctuates rapidly due to accretion, feedback, and stripping, the stellar component evolves more gradually and retains a record of past activity. SFH curves thus offer meaningful insight into the evolutionary path of a galaxy. Figure~\ref{fig:SFHall} shows normalized cumulative SFHs for Local Group dwarf galaxies from \cite{Weisz2014}, compared with subhalos in the five baryonic Mochima runs. The number of matched satellites varies between runs; for instance, in the \blue simulation, the high central concentration of the host halo leads to the disruption of many subhalos that survive in other realizations.

A dedicated study of the NewHorizon simulation \cite{Dubois2021NH} on the formation of dark matter cores reinforces the scenario described above, suggesting that cusp–core transformations are driven by repeated, bursty episodes of star formation. Supernova feedback expels central gas on timescales of approximately 2–3 Gyr, altering the gravitational potential and causing dark matter to migrate outward. However, if central star formation continues rapidly, the accumulated stellar mass can eventually dominate the gravitational potential, leading to adiabatic contraction and the reformation of a cusp \cite{Jackson2024}.

While this mechanism has been proposed previously, the NewHorizon study is particularly relevant because its subgrid physics setup lies between those adopted in the \rose and \rouge simulations, with the additional inclusion of AGN feedback at the centre. The Mochima simulations offer a complementary perspective by isolating the role of stellar feedback alone, allowing for a more focused examination of its effect on dark matter in the absence of AGN-driven processes. 

\begin{figure}[t]
\centering
\begin{subfigure}[b]{0.43\linewidth}
\includegraphics[width=\linewidth]{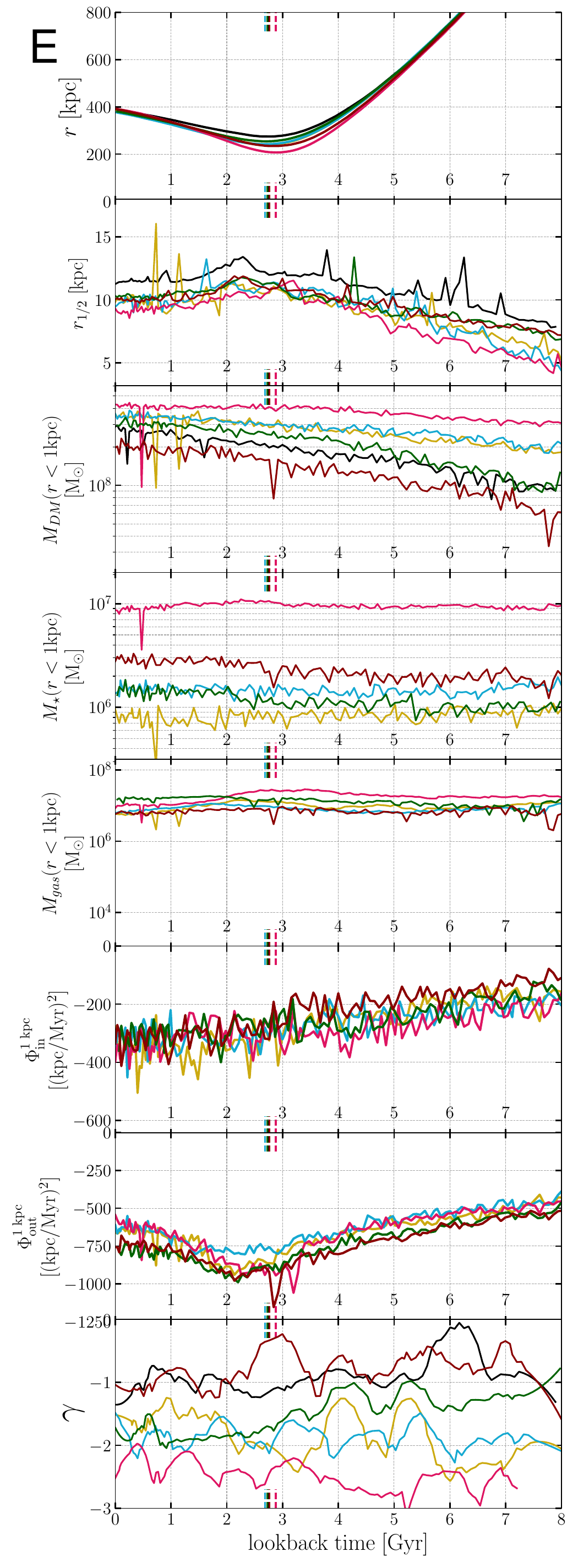}
\end{subfigure}
\begin{subfigure}[b]{0.43\linewidth}
\includegraphics[width=\linewidth]{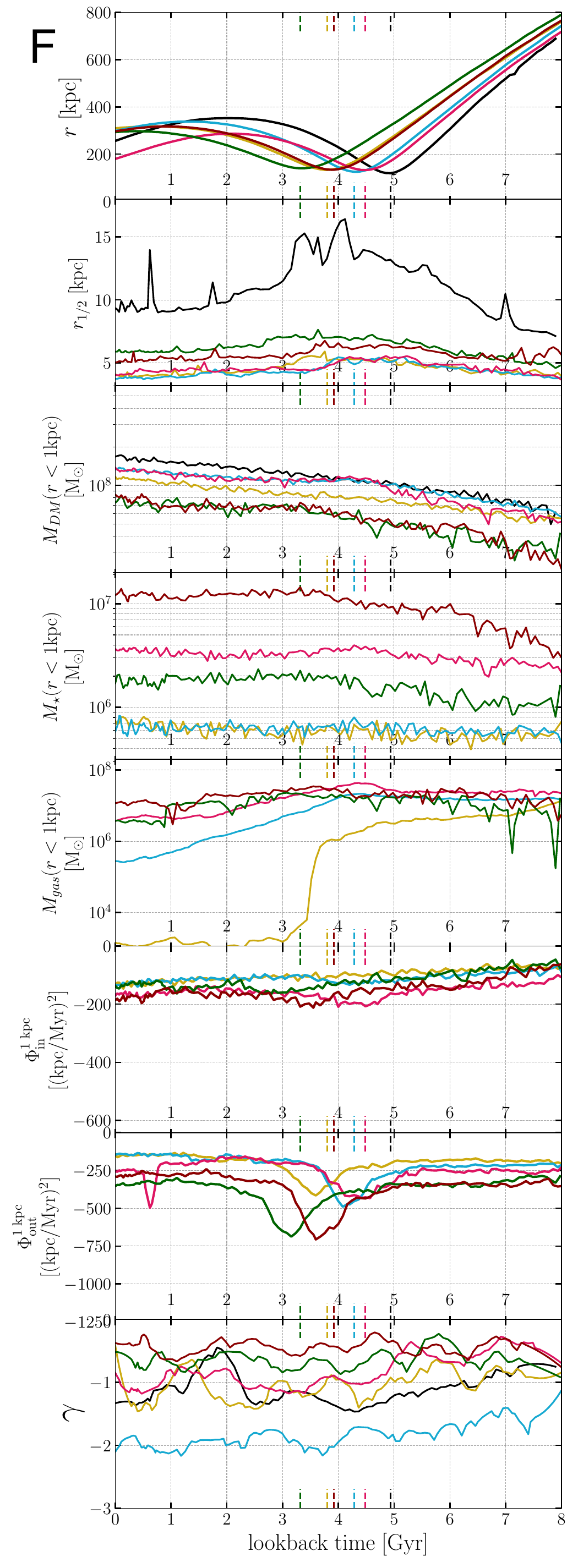}
\end{subfigure}
\caption{Time evolution of two representative subhalos from the Mochima simulations. Each subhalo is shown in the color corresponding to their simulation run. Each row shows a physical quantity as a function of lookback time: distance to the central galaxy, half-mass radius, enclosed dark matter mass, stellar mass, gas mass, internal and external components of the gravitational potential within the central 1 kpc, and the inner dark matter slope $\gamma$. Colored ticks indicate the time of pericentric passage.}
\label{fig:evo}%
\end{figure}
Figure~\ref{fig:SFHall} reveals a highly diverse set of SFH curves, even within individual simulations. This diversity suggests that a similarly wide range of inner dark matter structures is likely present across the satellite population. Building on the premise that stellar activity within the past 2–3 Gyr can significantly influence the inner dark matter profile, the SFH curves serve as useful proxies for understanding the evolving gravitational potential and its impact on the central dark matter distribution.

Let us first consider the case of a satellite galaxy with a relatively simple evolutionary history, experiencing minimal interactions with other halos. Key features of such a system are illustrated in the left set of panels in Figure~\ref{fig:evo}, this galaxy is tagged as halo E. All plots share a common horizontal axis representing lookback time, covering the past 8 Gyr. The top panel shows the distance of halo E from the main galaxy, with its closest approach occurring 3 Gyr ago at a distance of approximately 200~kpc, the full orbit is shown in figure \ref{fig:orbits}. The second panel shows the half-mass radius, $r_{1/2}$, defined as the radius that contains half of the total bounded DM mass. Then, the third, fourth, and fifth panels display the evolution of dark matter mass, stellar mass, and gas mass within 1kpc of its centre, respectively. The evolution of this subhalo is similar across all simulations. The half-mass radius grows as halo E approaches pericentric passage, after which its expansion stagnates or even reverses. In addition, the half-mass radius in the DMO run is larger than in the baryonic runs, likely due to halo contraction induced by the presence of the baryonic component. The stellar mass follows the hierarchy observed in the central galaxy, with runs including mechanical feedback producing stars more efficiently than those with delayed cooling, and runs with turbulent star formation outperforming the run with fixed star formation efficiency. The gas content remains relatively constant within the central region, while the dark matter mass increases steadily until the pericentric passage, after which its growth slows down significantly.

The fifth and sixth panels show the internal and external components of the gravitational potential, computed within a spherically symmetric mass distribution and defined as:
\begin{align}
\Phi(r) &= \Phi_{\rm in}(r) + \Phi_{\rm out}(r) \\
        &= -\,G \Biggl[
           \frac{M(r)}{r}
           + \int_{r}^{r_{\rm max}} \frac{4\pi\,r'^{2}\,\rho(r')}{r'} \,\mathrm{d}r'
         \Biggr].
\end{align}

\noindent Here, $G$ is the gravitational constant, $M(r)$ is the mass enclosed within radius $r$, $r_{\rm max}$ is set to $2r_{\rm vir}$, and $\rho(r)$ is the total mass density, including dark matter, stars, and gas. The evolution of both components of the potential remains stable, with only a minor dip in $\Phi_{\rm out}$ near the time of pericentric passage. As a result, the inner slope of the dark matter profile, $\gamma = \textrm{d}\ln \rho/\textrm{d}\ln r$, remains steep throughout as shown in the bottom of Figure \ref{fig:evo}. In all simulations except \rose, the central cusp forms shortly after the cessation of star formation, which occurs around 8 Gyr ago.

The case of halo E illustrates a quiet and uneventful evolutionary history. In contrast, the right set of panels in Figure~\ref{fig:evo} shows the evolution of halo F, a more dynamically active system across all simulations (see also Figure~\ref{fig:SFHall}). In all cases except \jaune, halo F continues forming stars until approximately 5~Myr ago.

The influence of the host potential on halo F varies significantly between runs, with the timing of pericentric passages differing across simulations. In all models, the central dark matter mass increases steadily, in some cases by nearly an order of magnitude. Although the stellar mass remains relatively stable after pericentric passage, differences between feedback models are evident: simulations employing Mechanical Feedback produce up to ten times more stellar mass within the central kiloparsec than those with Delayed Cooling. This stellar content then helps reduce the effect of ram stripping of the gas, as runs with more stars are able to keep most of their gas after pericentric passage while the two runs with  the lower stellar masses are stripped of most of their gas by the central galaxy. The absence of gas quenches star formation and its feedback, allowing the stability of the DM cusp. The post-pericentre behaviour is consistent with an impulsive tidal heating episode, in which rapid changes in the external potential inject energy into the bound component \cite{Jiang2019,ErraniPenarrubia2020}. This is indicated by the evolution of the half-mass radius, which shows that subhalo F in all runs, undergoes an expansion immediately after pericentric passage before subsequently contracting. This behaviour is clearer in the DMO run where the effect is more drastic. 

The central slope of the dark matter profiles, shown in the lowermost panels of Figure \ref{fig:evo} through the $\gamma$ parameter, is complex and likely linked to stellar mass and supernova feedback (see Section \ref{sec:Discussion}). To support the values of $\gamma$ discussed here, Figure \ref{fig:exampleprofiles} presents the evolution of the dark matter profiles for the two extreme cases of subhalos E and F. For halo E, a very steep cusp is observed in the \rose run, where the stellar mass exceeds that of the other runs by an order of magnitude already 8 Gyr ago, inducing contraction of the dark matter halo. At the opposite end of the $\gamma$ values, the \rouge run, also with mechanical feedback, produces a shallower cusp that resembles the behaviour of the DMO run. 

The case of halo F is more diverse. The steepest central profile is seen in the \blue run, which can be explained by steady stellar mass growth in the centre, with a star formation history that plateaued around 8 Gyr ago, allowing the halo to establish a robust cusp. At the other end of the $\gamma$ spectrum, cores are found in the two runs that exhibit late star formation, as shown in Figure \ref{fig:SFHall} and by the continued growth of the stellar mass within 1 kpc until the pericentric passage around 4 Gyr ago. In these runs, the gas mass within 1 kpc also shows larger fluctuations than in other cases, likely driven by supernova feedback from recently formed stars. Such fluctuations in the gas component have been invoked to explain core formation in dwarf galaxies, as discussed in Section \ref{sec:Discussion}. However, the cases with cores ($\gamma > -1$) are not stable: $\gamma$ oscillates between very flat values (cores $\gamma >-1$) and values close to unity. This behaviour is further illustrated in Figure \ref{fig:exampleprofiles}, where profiles with cusps show less dispersion over time than profiles with cores.

The evolution of halo F illustrates the range of physical processes that shape the central dark matter structure in dwarf galaxies. Sustained star formation drives episodic gas outflows and mass redistribution, while interactions with the host modify the external gravitational field and influence the internal gas content. Together, these effects produce temporary core–cusp transitions over Gyr timescales. In contrast, more quiescent systems such as halo E remain stable over long periods, preserving a central cusp.

\begin{figure}[t]
\centering
\begin{subfigure}[b]{0.98\linewidth}
\includegraphics[width=\linewidth]{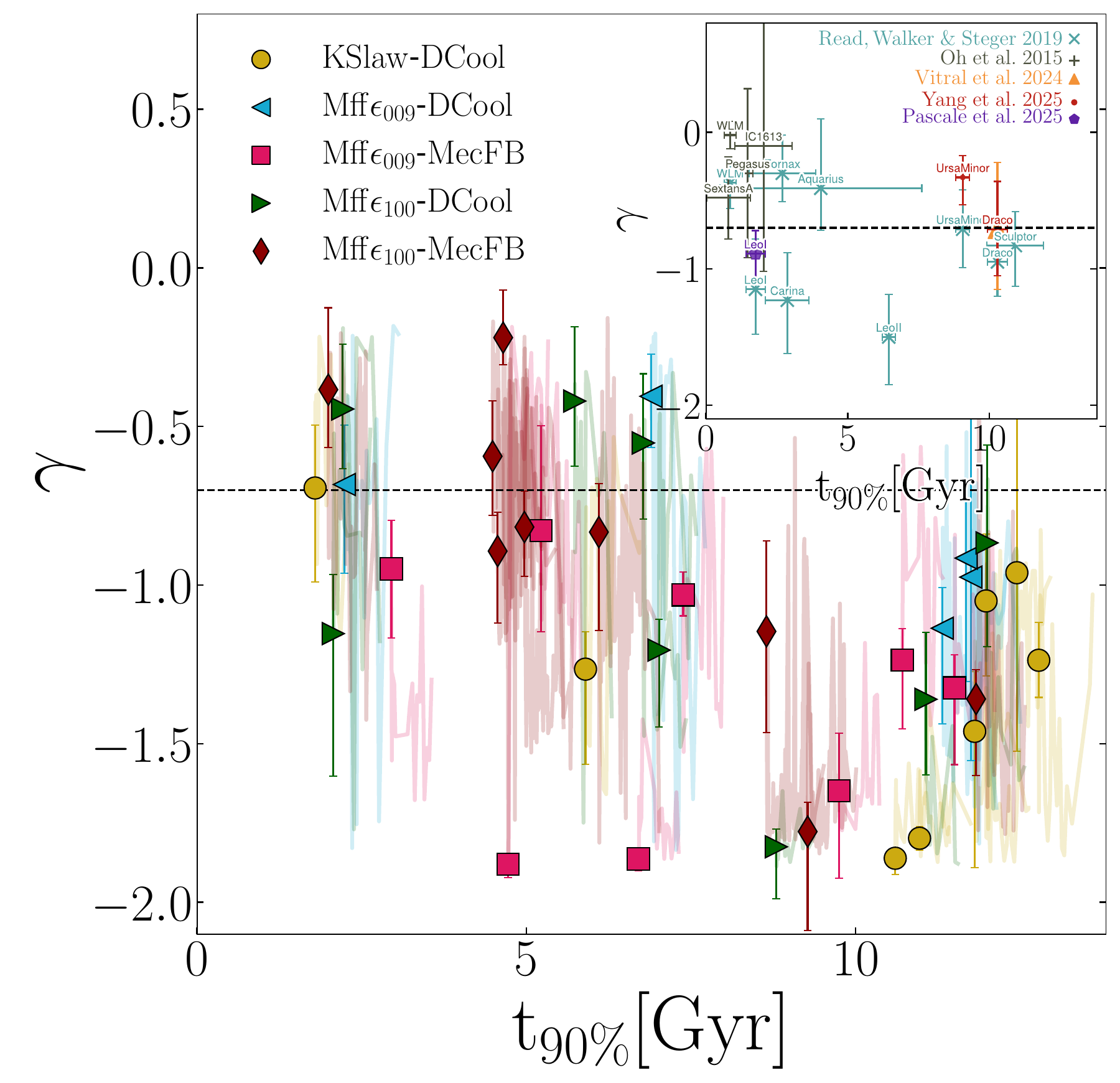}
\end{subfigure}
\caption{The relation between the inner dark matter slope, $\gamma$, and the time at which a dwarf galaxy forms 90$\%$ of its total stellar mass, $t_{90\%}$. Dwarf satellite galaxies from the simulations, showing the recovered evolution of $\gamma$ over the last 8 Gyr, compressed into a 1 Gyr timescale as a shaded line connected to each simulation point. The inner panel shows predictions for dwarf galaxies in the Local Group, where inner slopes are taken from various studies referenced in the image, and the $t_{90\%}$ parameter is derived from the SFH curves presented in Weisz et al. 2014 \cite{Weisz2014}.}
\label{fig:T90plot}%
\end{figure}
\section{Discussion}\label{sec:Discussion}

The results described above suggest the existence of at least two distinct regimes in the relationship between a galaxy star formation history (SFH) and the inner slope of its dark matter profile, $\gamma$.

The first regime corresponds to systems that form the bulk of their stellar mass early in their evolution. Once star formation ceases or stagnates, the central gravitational potential stabilizes, allowing a cusp to persist or to grow further through contraction. The second regime is dynamically active and includes galaxies that continue forming stars until late cosmic times. In these systems, recent stellar feedback—particularly from supernova-driven gas outflows—can induce repeated fluctuations in the central gravitational potential, leading to alternating phases of core formation and re-contraction. The resulting $\gamma$ becomes unstable and may vary significantly over time.

In other words, galaxies that formed 90$\%$ of their stellar mass more than 7~Gyr ago typically show a stable cusp at redshift zero. By contrast, galaxies where more than 10$\%$ of the stellar mass was assembled within the last 5~Gyr are more likely to reside in the fluctuating regime, where both core-like and cusp-like profiles can appear depending on the timing of recent feedback events.

Figure~\ref{fig:T90plot} illustrates this dual behaviour. The inner panel shows observational data for Local Group dwarf galaxies, where inner slopes are taken from \cite{Read2019,Oh2015,Vitral2024,Pascale2025,Yang2025}, and the corresponding $t_{90\%}$ values are derived from the cumulative SFH curves of \cite{Weisz2014}, shown in the upper left panel of Figure~\ref{fig:SFHall}\footnote{Horizontal error bars on $t_{90\%}$ are computed using linear error propagation: $\Delta t = \Delta \textrm{SFH} / \frac{d\,(\textrm{SFH})}{dt}$, assuming a 10$\%$ uncertainty on the SFH curve. Note that Draco and WLM appear twice in this panel, based on different data sources.}.

The main panel presents the corresponding relation for subhalos in the simulations. Here, $t_{90\%}$ is measured precisely, and vertical error bars on $\gamma$ correspond to the uncertainty from MCMC fitting (see Paper~2 for methodological details). The shaded lines extending from each simulation point depict the full evolutionary trajectory of $\gamma$ over the last 8~Gyr, compressed horizontally into a 1~Gyr timespan in $t_{90\%}$ space, thereby illustrating the dynamical evolution of the inner dark matter slope in each subhalo.

The scenario is consistent across both observational and simulated samples. Galaxies with $t_{90\%} \gtrsim 7$~Gyr invariably exhibit cusps profiles and show minimal evolution in $\gamma$ over time. In contrast, galaxies with $t_{90\%} \lesssim 7$~Gyr show a wide spread in $\gamma$ and are characterized by significant temporal fluctuations. This suggests that early truncation of star formation is a key factor in stabilizing cusps, whereas extended or bursty late-time star formation places galaxies into a regime of ongoing dynamical transformation of their central dark matter distribution.

There are already several hints in the literature pointing in the direction proposed here. For instance, the dependence of the inner dark matter profile on the star formation subgrid model was demonstrated in a study exploring variations in the gas density threshold for star formation, under the assumption of a fixed star formation efficiency \cite{Dutton:2018nop}. Using a setup comparable to the \jaune\ run of the Mochima suite, this study showed that when the star formation threshold is too low (e.g., $n \sim 0.1$~cm$^{-3}$), star formation proceeds too efficiently, preventing the formation of cores in the dark matter profile.

At much higher resolution, the EDGE simulations have revealed a strong correlation between the timing of star formation and the shape of the inner dark matter profile for isolated dwarf galaxies in a similar mass range, even showing recontraction of the DM halos by late mass accretions \cite{Orkey2021,Muni2025}. In that study, systems where most stars formed early tended to develop cusps, while galaxies with more extended star formation histories formed cores. Because the EDGE sample includes only field dwarfs, the observed "diversity" in $\gamma$ can be attributed almost entirely to differences in star formation timing. In contrast, the Mochima simulations suggest that satellites of similar mass are subject to additional complexity: gravitational interactions with the host galaxy and neighboring satellites can induce oscillations in the central dark matter density over time. This behaviour is clearly seen in the $\gamma$-evolution tracks of Figure~\ref{fig:T90plot}, where satellites with late star formation histories exhibit more variable inner slopes.
\smallskip
\noindent
One caveat to this analysis is that the structural response of subhalos to tidal effects can be sensitive to numerical resolution. As shown by \cite{Borukhovetskaya2022}, insufficient resolution can lead to artificially inflated remnant sizes and deviations from expected tidal evolution tracks, particularly when only a small number of particles remain bound. While the Mochima simulations resolve massive subhalos with adequate particle number, caution is warranted when interpreting the outcome of extreme tidal stripping, especially for low-mass or heavily stripped systems.

\section{Summary - Conclusions}\label{sec:summary}

This study has presented a detailed analysis of the subhalo population in the suite of Mochima simulations, focusing on the influence of baryonic physics on subhalo survival, stellar content, and inner dark matter structure. A refined method was developed to identify bound subhalo material by isolating the gravitational potential of each structure and applying phase-space selection criteria. This approach reveals trends that depend not only on the depth of the central gravitational potential but also on its shape and evolution, both of which are modulated by the baryonic physics models implemented.

At the population level, the number and structure of surviving subhalos are governed primarily by the depth of the central gravitational potential and the internal structure of the subhalos themselves. The structure of the host dark matter halo also influences the survival of low-mass satellites: hosts with lower concentration, linked to the absence of protostellar feedback, exhibit shallower outer potentials and retain larger populations of low-mass subhalos. In contrast, runs with deeper central potentials show stronger subhalo depletion, particularly at intermediate and high masses. At the high-mass end ($M > 5 \times 10^8$~M$_\odot$), the survivability of satellites is further enhanced by their baryonic content, as the buildup of internal stellar mass deepens their gravitational potentials and improves their resistance to disruption. Overall, the present simulations do not show strong signs of the ``too big to fail'' problem, as the stellar mass spectrum of subhalos is consistent with observations of the MW and M31 disc galaxies.

The inner dark matter slope of individual subhalos shows substantial diversity, consistent with observational studies of rotation curves, and is shaped by a combination of internal stellar feedback and environmental interactions, such as ram-pressure stripping and tidal heating. By tracing the star formation history and gravitational potential evolution of resolved satellites, this study demonstrates that both quiescent and dynamically active evolutionary tracks are possible. Systems that complete most of their star formation early tend to retain stable cusps, while those with extended or recent star formation frequently exhibit fluctuating inner slopes driven by bursty feedback and tidal perturbations. A clear connection is identified between the epoch of stellar mass assembly ($t_{90\%}$) and the present-day inner dark matter profile, in agreement with both simulation predictions and observational constraints. The added influence of external tides further impacts the long-term stability of the central dark matter profile, complicating the picture. These results suggest that the evolution of inner slopes is a complex and time-dependent process, challenging simplified analytical approaches. Comparisons with such models will be pursued in a subsequent study.

Among the different runs of the Mochima simulation, the satellites from the two extreme cases in stellar mass - \jaune\ with the lowest star formation and \rose\ with the highest - both exhibit central cusps, with cores being rare. In the low stellar mass case, feedback is insufficient to modify the inner dark matter distribution, producing profiles that closely resemble those of DMO halos ($\gamma \sim -1$). In contrast, the high stellar mass content of \rose, produced by an early and efficient star formation history, leads to a steep and monotonic stellar buildup that suppresses the bursty feedback episodes required for core formation inducing steeper cusps($\gamma< -1$). The remaining three simulations, which benefit from more balanced baryonic configurations, produce a diversity of subhalo structures consistent with observations. In these cases, core formation is likely enabled by recent stellar activity occurring within shallow gravitational wells, unlike scenarios with ultra-efficient early star formation where core formation is suppressed.

Taken together, these results indicate that clump properties and inner structure are regulated by the combined effects of multi-scale feedback mechanisms and environmental processes. The observed diversity in inner dark matter structure -often viewed as a challenge to cold dark matter models- can arise naturally from the interplay between star formation history and environmental context. Future work will extend this analysis to trace substructure evolution across cosmic time and to investigate the survival and transformation of tidal features before redshift zero.

\acknowledgments

The authors thank Jean-Charles Lambert for his valuable support, without which this research would have been significantly more difficult. We would also like to thank Benoit Famaey and Jonathan Freundlich for their insightful and enriching discussions regarding the evolution of dark matter halos in the present context. 
This research has made use of computing facilities operated by CeSAM(1) data centre at LAM(2), Marseille, France
(1) CeSAM : Centre de données Astrophysiques de Marseille
(2) LAM : Laboratoire Astrophysique de Marseille
This work has been carried out thanks to the support of the OCEVU Labex (ANR-11-LABX-0060) and the
A*MIDEX project (ANR-11-IDEX-0001-02) funded by the Investissements d'Avenir French government
program managed by the ANR.
This work was granted access to the HPC resources of Aix-Marseille Universit\'e financed by the project Equip@Meso (ANR-10-EQPX-29-01) of the program  ``Investissements d'Avenir'' supervised by the Agence Nationale de la Recherche.
This work has been supported by funding from the ANR project ANR-18-CE31-0006 (GaDaMa)

 \bibliography{subhalos} 

\providecommand{\href}[2]{#2}\begingroup\raggedright\begin{thebibliography}{100}

\bibitem{WhiteRees1978}
S.~D.~M. {White} and M.~J. {Rees}, \emph{{Core condensation in heavy halos: a
  two-stage theory for galaxy formation and clustering.}},
  \href{https://doi.org/10.1093/mnras/183.3.341}{\emph{\mnras} {\bfseries 183}
  (1978) 341}.

\bibitem{WechslerTinker2018}
R.~H. {Wechsler} and J.~L. {Tinker}, \emph{{The Connection Between Galaxies and
  Their Dark Matter Halos}},
  \href{https://doi.org/10.1146/annurev-astro-081817-051756}{\emph{\araa}
  {\bfseries 56} (2018) 435}
  [\href{https://arxiv.org/abs/1804.03097}{{\ttfamily 1804.03097}}].

\bibitem{Hayashi2003}
E.~{Hayashi}, J.~F. {Navarro}, J.~E. {Taylor}, J.~{Stadel} and T.~{Quinn},
  \emph{{The Structural Evolution of Substructure}},
  \href{https://doi.org/10.1086/345788}{\emph{\apj} {\bfseries 584} (2003) 541}
  [\href{https://arxiv.org/abs/astro-ph/0203004}{{\ttfamily
  astro-ph/0203004}}].

\bibitem{Moore2000}
B.~{Moore}, S.~{Gelato}, A.~{Jenkins}, F.~R. {Pearce} and V.~{Quilis},
  \emph{{Collisional versus Collisionless Dark Matter}},
  \href{https://doi.org/10.1086/312692}{\emph{\apjl} {\bfseries 535} (2000)
  L21} [\href{https://arxiv.org/abs/astro-ph/0002308}{{\ttfamily
  astro-ph/0002308}}].

\bibitem{Onions2012}
J.~{Onions}, A.~{Knebe}, F.~R. {Pearce}, S.~I. {Muldrew}, H.~{Lux}, S.~R.
  {Knollmann} et~al., \emph{{Subhaloes going Notts: the subhalo-finder
  comparison project}},
  \href{https://doi.org/10.1111/j.1365-2966.2012.20947.x}{\emph{\mnras}
  {\bfseries 423} (2012) 1200}
  [\href{https://arxiv.org/abs/1203.3695}{{\ttfamily 1203.3695}}].

\bibitem{Mansfield2024}
P.~{Mansfield}, E.~{Darragh-Ford}, Y.~{Wang}, E.~O. {Nadler}, B.~{Diemer} and
  R.~H. {Wechsler}, \emph{{SYMFIND : Addressing the Fragility of Subhalo
  Finders and Revealing the Durability of Subhalos}},
  \href{https://doi.org/10.3847/1538-4357/ad4e33}{\emph{\apj} {\bfseries 970}
  (2024) 178} [\href{https://arxiv.org/abs/2308.10926}{{\ttfamily
  2308.10926}}].

\bibitem{Pathak2025}
D.~{Pathak}, C.~R. {Christensen}, A.~M. {Brooks}, F.~{Munshi}, A.~C. {Wright}
  and C.~{Carter}, \emph{{Survivors and Zombies: The Quenching and Disruption
  of Satellites around Milky Way Analogs}},
  \href{https://doi.org/10.48550/arXiv.2505.22742}{\emph{arXiv e-prints} (2025)
  arXiv:2505.22742} [\href{https://arxiv.org/abs/2505.22742}{{\ttfamily
  2505.22742}}].

\bibitem{Boylan-Kolchin2011}
M.~{Boylan-Kolchin}, J.~S. {Bullock} and M.~{Kaplinghat}, \emph{{Too big to
  fail? The puzzling darkness of massive Milky Way subhaloes}},
  \href{https://doi.org/10.1111/j.1745-3933.2011.01074.x}{\emph{\mnras}
  {\bfseries 415} (2011) L40}
  [\href{https://arxiv.org/abs/1103.0007}{{\ttfamily 1103.0007}}].

\bibitem{deBlok2008}
W.~J.~G. {de Blok}, F.~{Walter}, E.~{Brinks}, C.~{Trachternach}, S.~H. {Oh} and
  R.~C. {Kennicutt}, Jr., \emph{{High-Resolution Rotation Curves and Galaxy
  Mass Models from THINGS}},
  \href{https://doi.org/10.1088/0004-6256/136/6/2648}{\emph{\aj} {\bfseries
  136} (2008) 2648} [\href{https://arxiv.org/abs/0810.2100}{{\ttfamily
  0810.2100}}].

\bibitem{Oh2011}
S.-H. {Oh}, C.~{Brook}, F.~{Governato}, E.~{Brinks}, L.~{Mayer}, W.~J.~G. {de
  Blok} et~al., \emph{{The Central Slope of Dark Matter Cores in Dwarf
  Galaxies: Simulations versus THINGS}},
  \href{https://doi.org/10.1088/0004-6256/142/1/24}{\emph{\aj} {\bfseries 142}
  (2011) 24} [\href{https://arxiv.org/abs/1011.2777}{{\ttfamily 1011.2777}}].

\bibitem{Moore1994}
B.~{Moore}, \emph{{Evidence against dissipation-less dark matter from
  observations of galaxy haloes}},
  \href{https://doi.org/10.1038/370629a0}{\emph{\nat} {\bfseries 370} (1994)
  629}.

\bibitem{Flores1994}
R.~A. {Flores} and J.~R. {Primack}, \emph{{Observational and Theoretical
  Constraints on Singular Dark Matter Halos}},
  \href{https://doi.org/10.1086/187350}{\emph{\apjl} {\bfseries 427} (1994) L1}
  [\href{https://arxiv.org/abs/astro-ph/9402004}{{\ttfamily
  astro-ph/9402004}}].

\bibitem{DelPopolo2021}
A.~{Del Popolo} and M.~{Le Delliou}, \emph{{Review of Solutions to the
  Cusp-Core Problem of the {\ensuremath{\Lambda}}CDM Model}},
  \href{https://doi.org/10.3390/galaxies9040123}{\emph{Galaxies} {\bfseries 9}
  (2021) 123} [\href{https://arxiv.org/abs/2209.14151}{{\ttfamily
  2209.14151}}].

\bibitem{Oman2015}
K.~A. {Oman}, J.~F. {Navarro}, A.~{Fattahi}, C.~S. {Frenk}, T.~{Sawala},
  S.~D.~M. {White} et~al., \emph{{The unexpected diversity of dwarf galaxy
  rotation curves}}, \href{https://doi.org/10.1093/mnras/stv1504}{\emph{\mnras}
  {\bfseries 452} (2015) 3650}
  [\href{https://arxiv.org/abs/1504.01437}{{\ttfamily 1504.01437}}].

\bibitem{Zolotov2012}
A.~{Zolotov}, A.~M. {Brooks}, B.~{Willman}, F.~{Governato}, A.~{Pontzen},
  C.~{Christensen} et~al., \emph{{Baryons Matter: Why Luminous Satellite
  Galaxies have Reduced Central Masses}},
  \href{https://doi.org/10.1088/0004-637X/761/1/71}{\emph{\apj} {\bfseries 761}
  (2012) 71} [\href{https://arxiv.org/abs/1207.0007}{{\ttfamily 1207.0007}}].

\bibitem{DelPopolo2014}
A.~{Del Popolo}, J.~A.~S. {Lima}, J.~C. {Fabris} and D.~C. {Rodrigues},
  \emph{{A unified solution to the small scale problems of the
  {\ensuremath{\Lambda}}CDM model}},
  \href{https://doi.org/10.1088/1475-7516/2014/04/021}{\emph{\jcap} {\bfseries
  2014} (2014) 021} [\href{https://arxiv.org/abs/1404.3674}{{\ttfamily
  1404.3674}}].

\bibitem{Mashchenko2006}
S.~{Mashchenko}, H.~M.~P. {Couchman} and J.~{Wadsley}, \emph{{The removal of
  cusps from galaxy centres by stellar feedback in the early Universe}},
  \href{https://doi.org/10.1038/nature04944}{\emph{\nat} {\bfseries 442} (2006)
  539} [\href{https://arxiv.org/abs/astro-ph/0605672}{{\ttfamily
  astro-ph/0605672}}].

\bibitem{Mashchenko2008}
S.~{Mashchenko}, J.~{Wadsley} and H.~M.~P. {Couchman}, \emph{{Stellar Feedback
  in Dwarf Galaxy Formation}},
  \href{https://doi.org/10.1126/science.1148666}{\emph{Science} {\bfseries 319}
  (2008) 174} [\href{https://arxiv.org/abs/0711.4803}{{\ttfamily 0711.4803}}].

\bibitem{Penarrubia2010}
J.~{Pe{\~n}arrubia}, A.~J. {Benson}, M.~G. {Walker}, G.~{Gilmore}, A.~W.
  {McConnachie} and L.~{Mayer}, \emph{{The impact of dark matter cusps and
  cores on the satellite galaxy population around spiral galaxies}},
  \href{https://doi.org/10.1111/j.1365-2966.2010.16762.x}{\emph{\mnras}
  {\bfseries 406} (2010) 1290}
  [\href{https://arxiv.org/abs/1002.3376}{{\ttfamily 1002.3376}}].

\bibitem{Errani2023}
R.~{Errani}, J.~F. {Navarro}, J.~{Pe{\~n}arrubia}, B.~{Famaey} and R.~{Ibata},
  \emph{{Dark matter halo cores and the tidal survival of Milky Way
  satellites}}, \href{https://doi.org/10.1093/mnras/stac3499}{\emph{\mnras}
  {\bfseries 519} (2023) 384}
  [\href{https://arxiv.org/abs/2210.01131}{{\ttfamily 2210.01131}}].

\bibitem{Celiz2025}
B.~M. {Celiz}, J.~F. {Navarro}, M.~G. {Abadi} and V.~{Springel},
  \emph{{Mass-morphology relation of TNG50 galaxies}},
  \href{https://doi.org/10.1051/0004-6361/202554847}{\emph{\aap} {\bfseries
  699} (2025) A12} [\href{https://arxiv.org/abs/2505.01620}{{\ttfamily
  2505.01620}}].

\bibitem{Sommer-Larsen2001}
J.~{Sommer-Larsen} and A.~{Dolgov}, \emph{{Formation of Disk Galaxies: Warm
  Dark Matter and the Angular Momentum Problem}},
  \href{https://doi.org/10.1086/320211}{\emph{\apj} {\bfseries 551} (2001) 608}
  [\href{https://arxiv.org/abs/astro-ph/9912166}{{\ttfamily
  astro-ph/9912166}}].

\bibitem{Peebles2000}
P.~J.~E. {Peebles}, \emph{{Fluid Dark Matter}},
  \href{https://doi.org/10.1086/312677}{\emph{\apjl} {\bfseries 534} (2000)
  L127} [\href{https://arxiv.org/abs/astro-ph/0002495}{{\ttfamily
  astro-ph/0002495}}].

\bibitem{Goodman2000}
J.~{Goodman}, \emph{{Repulsive dark matter}},
  \href{https://doi.org/10.1016/S1384-1076(00)00015-4}{\emph{na} {\bfseries 5}
  (2000) 103} [\href{https://arxiv.org/abs/astro-ph/0003018}{{\ttfamily
  astro-ph/0003018}}].

\bibitem{Hu2000}
W.~{Hu}, R.~{Barkana} and A.~{Gruzinov}, \emph{{Fuzzy Cold Dark Matter: The
  Wave Properties of Ultralight Particles}},
  \href{https://doi.org/10.1103/PhysRevLett.85.1158}{\emph{\prl} {\bfseries 85}
  (2000) 1158} [\href{https://arxiv.org/abs/astro-ph/0003365}{{\ttfamily
  astro-ph/0003365}}].

\bibitem{Cen2001}
R.~{Cen}, \emph{{Decaying Cold Dark Matter Model and Small-Scale Power}},
  \href{https://doi.org/10.1086/318861}{\emph{\apjl} {\bfseries 546} (2001)
  L77} [\href{https://arxiv.org/abs/astro-ph/0005206}{{\ttfamily
  astro-ph/0005206}}].

\bibitem{Kaplinghat2000}
M.~{Kaplinghat}, L.~{Knox} and M.~S. {Turner}, \emph{{Annihilating Cold Dark
  Matter}}, \href{https://doi.org/10.1103/PhysRevLett.85.3335}{\emph{\prl}
  {\bfseries 85} (2000) 3335}
  [\href{https://arxiv.org/abs/astro-ph/0005210}{{\ttfamily
  astro-ph/0005210}}].

\bibitem{Spergel2000}
D.~N. {Spergel} and P.~J. {Steinhardt}, \emph{{Observational Evidence for
  Self-Interacting Cold Dark Matter}},
  \href{https://doi.org/10.1103/PhysRevLett.84.3760}{\emph{\prl} {\bfseries 84}
  (2000) 3760} [\href{https://arxiv.org/abs/astro-ph/9909386}{{\ttfamily
  astro-ph/9909386}}].

\bibitem{Zentner2003}
A.~R. {Zentner} and J.~S. {Bullock}, \emph{{Halo Substructure and the Power
  Spectrum}}, \href{https://doi.org/10.1086/378797}{\emph{\apj} {\bfseries 598}
  (2003) 49} [\href{https://arxiv.org/abs/astro-ph/0304292}{{\ttfamily
  astro-ph/0304292}}].

\bibitem{Milgrom1983a}
M.~{Milgrom}, \emph{{A modification of the Newtonian dynamics - Implications
  for galaxies.}}, \href{https://doi.org/10.1086/161131}{\emph{\apj} {\bfseries
  270} (1983) 371}.

\bibitem{Milgrom1983b}
M.~{Milgrom}, \emph{{A modification of the Newtonian dynamics as a possible
  alternative to the hidden mass hypothesis.}},
  \href{https://doi.org/10.1086/161130}{\emph{\apj} {\bfseries 270} (1983)
  365}.

\bibitem{Buchdahl1970}
H.~A. {Buchdahl}, \emph{{Non-linear Lagrangians and cosmological theory}},
  \href{https://doi.org/10.1093/mnras/150.1.1}{\emph{\mnras} {\bfseries 150}
  (1970) 1}.

\bibitem{Starobinsky1980}
A.~A. {Starobinsky}, \emph{{A new type of isotropic cosmological models without
  singularity}},
  \href{https://doi.org/10.1016/0370-2693(80)90670-X}{\emph{Physics Letters B}
  {\bfseries 91} (1980) 99}.

\bibitem{Bengochea2009}
G.~R. {Bengochea} and R.~{Ferraro}, \emph{{Dark torsion as the cosmic
  speed-up}}, \href{https://doi.org/10.1103/PhysRevD.79.124019}{\emph{\prd}
  {\bfseries 79} (2009) 124019}
  [\href{https://arxiv.org/abs/0812.1205}{{\ttfamily 0812.1205}}].

\bibitem{Dent2011}
J.~B. {Dent}, S.~{Dutta} and E.~N. {Saridakis}, \emph{{f(T) gravity mimicking
  dynamical dark energy. Background and perturbation analysis}},
  \href{https://doi.org/10.1088/1475-7516/2011/01/009}{\emph{\jcap} {\bfseries
  2011} (2011) 009} [\href{https://arxiv.org/abs/1010.2215}{{\ttfamily
  1010.2215}}].

\bibitem{deBlok2001}
W.~J.~G. {de Blok}, S.~S. {McGaugh}, A.~{Bosma} and V.~C. {Rubin}, \emph{{Mass
  Density Profiles of Low Surface Brightness Galaxies}},
  \href{https://doi.org/10.1086/320262}{\emph{\apjl} {\bfseries 552} (2001)
  L23} [\href{https://arxiv.org/abs/astro-ph/0103102}{{\ttfamily
  astro-ph/0103102}}].

\bibitem{deBlok2003}
W.~J.~G. {de Blok}, A.~{Bosma} and S.~{McGaugh}, \emph{{Simulating observations
  of dark matter dominated galaxies: towards the optimal halo profile}},
  \href{https://doi.org/10.1046/j.1365-8711.2003.06330.x}{\emph{\mnras}
  {\bfseries 340} (2003) 657}
  [\href{https://arxiv.org/abs/astro-ph/0212102}{{\ttfamily
  astro-ph/0212102}}].

\bibitem{Power2002}
C.~Power, J.~Navarro, A.~Jenkins, C.~Frenk, S.~D. White, V.~Springel et~al.,
  \emph{{The Inner structure of Lambda CDM halos. 1. A Numerical convergence
  study}}, \href{https://doi.org/10.1046/j.1365-8711.2003.05925.x}{\emph{Mon.
  Not. Roy. Astron. Soc.} {\bfseries 338} (2003) 14}
  [\href{https://arxiv.org/abs/astro-ph/0201544}{{\ttfamily
  astro-ph/0201544}}].

\bibitem{DiCintio:2013qxa}
A.~Di~Cintio, C.~B. Brook, A.~V. Macci\`o, G.~S. Stinson, A.~Knebe, A.~A.
  Dutton et~al., \emph{{The dependence of dark matter profiles on the
  stellar-to-halo mass ratio: a prediction for cusps versus cores}},
  \href{https://doi.org/10.1093/mnras/stt1891}{\emph{Mon. Not. Roy. Astron.
  Soc.} {\bfseries 437} (2014) 415}
  [\href{https://arxiv.org/abs/1306.0898}{{\ttfamily 1306.0898}}].

\bibitem{Tollet2016}
E.~{Tollet}, A.~V. {Macci{\`o}}, A.~A. {Dutton}, G.~S. {Stinson}, L.~{Wang},
  C.~{Penzo} et~al., \emph{{NIHAO - IV: core creation and destruction in dark
  matter density profiles across cosmic time}},
  \href{https://doi.org/10.1093/mnras/stv2856}{\emph{MNRAS} {\bfseries 456}
  (2016) 3542} [\href{https://arxiv.org/abs/1507.03590}{{\ttfamily
  1507.03590}}].

\bibitem{Chan2015}
T.~K. {Chan}, D.~{Kere{\v{s}}}, J.~{O{\~n}orbe}, P.~F. {Hopkins}, A.~L.
  {Muratov}, C.~A. {Faucher-Gigu{\`e}re} et~al., \emph{{The impact of baryonic
  physics on the structure of dark matter haloes: the view from the FIRE
  cosmological simulations}},
  \href{https://doi.org/10.1093/mnras/stv2165}{\emph{\mnras} {\bfseries 454}
  (2015) 2981} [\href{https://arxiv.org/abs/1507.02282}{{\ttfamily
  1507.02282}}].

\bibitem{Jackson2024}
R.~A. {Jackson}, S.~{Kaviraj}, S.~K. {Yi}, S.~{Peirani}, Y.~{Dubois},
  G.~{Martin} et~al., \emph{{The formation of cores in galaxies across cosmic
  time - the existence of cores is not in tension with the
  {\ensuremath{\Lambda}}CDM paradigm}},
  \href{https://doi.org/10.1093/mnras/stae056}{\emph{\mnras} {\bfseries 528}
  (2024) 1655} [\href{https://arxiv.org/abs/2310.13055}{{\ttfamily
  2310.13055}}].

\bibitem{Read2016}
J.~I. {Read}, O.~{Agertz} and M.~L.~M. {Collins}, \emph{{Dark matter cores all
  the way down}}, \href{https://doi.org/10.1093/mnras/stw713}{\emph{\mnras}
  {\bfseries 459} (2016) 2573}
  [\href{https://arxiv.org/abs/1508.04143}{{\ttfamily 1508.04143}}].

\bibitem{Bose2019}
S.~{Bose}, C.~S. {Frenk}, A.~{Jenkins}, A.~{Fattahi}, F.~A. {G{\'o}mez},
  R.~J.~J. {Grand} et~al., \emph{{No cores in dark matter-dominated dwarf
  galaxies with bursty star formation histories}},
  \href{https://doi.org/10.1093/mnras/stz1168}{\emph{\mnras} {\bfseries 486}
  (2019) 4790} [\href{https://arxiv.org/abs/1810.03635}{{\ttfamily
  1810.03635}}].

\bibitem{Richings2020}
J.~{Richings}, C.~{Frenk}, A.~{Jenkins}, A.~{Robertson}, A.~{Fattahi}, R.~J.~J.
  {Grand} et~al., \emph{{Subhalo destruction in the APOSTLE and AURIGA
  simulations}}, \href{https://doi.org/10.1093/mnras/stz3448}{\emph{\mnras}
  {\bfseries 492} (2020) 5780}
  [\href{https://arxiv.org/abs/1811.12437}{{\ttfamily 1811.12437}}].

\bibitem{Nunez-Castineyra:2020ufe}
A.~{Nu{\~n}ez-Casti{\~n}eyra}, E.~{Nezri}, J.~{Devriendt} and R.~{Teyssier},
  \emph{{Cosmological simulations of the same spiral galaxy: the impact of
  baryonic physics}}, {\emph{arXiv e-prints} (2020) arXiv:2004.06008}
  [\href{https://arxiv.org/abs/2004.06008}{{\ttfamily 2004.06008}}].

\bibitem{Tollerud2008}
E.~J. {Tollerud}, J.~S. {Bullock}, L.~E. {Strigari} and B.~{Willman},
  \emph{{Hundreds of Milky Way Satellites? Luminosity Bias in the Satellite
  Luminosity Function}}, \href{https://doi.org/10.1086/592102}{\emph{\apj}
  {\bfseries 688} (2008) 277}
  [\href{https://arxiv.org/abs/0806.4381}{{\ttfamily 0806.4381}}].

\bibitem{McConnachie2012}
A.~W. {McConnachie}, \emph{{The Observed Properties of Dwarf Galaxies in and
  around the Local Group}},
  \href{https://doi.org/10.1088/0004-6256/144/1/4}{\emph{\aj} {\bfseries 144}
  (2012) 4} [\href{https://arxiv.org/abs/1204.1562}{{\ttfamily 1204.1562}}].

\bibitem{Fattahi2016}
A.~{Fattahi}, J.~F. {Navarro}, T.~{Sawala}, C.~S. {Frenk}, L.~V. {Sales},
  K.~{Oman} et~al., \emph{{The cold dark matter content of Galactic dwarf
  spheroidals: no cores, no failures, no problem}},
  \href{https://doi.org/10.48550/arXiv.1607.06479}{\emph{arXiv e-prints} (2016)
  arXiv:1607.06479} [\href{https://arxiv.org/abs/1607.06479}{{\ttfamily
  1607.06479}}].

\bibitem{Dubois2021NH}
Y.~{Dubois}, R.~{Beckmann}, F.~{Bournaud}, H.~{Choi}, J.~{Devriendt},
  R.~{Jackson} et~al., \emph{{Introducing the NEWHORIZON simulation: Galaxy
  properties with resolved internal dynamics across cosmic time}},
  \href{https://doi.org/10.1051/0004-6361/202039429}{\emph{\aap} {\bfseries
  651} (2021) A109} [\href{https://arxiv.org/abs/2009.10578}{{\ttfamily
  2009.10578}}].

\bibitem{Grand2017}
R.~J.~J. {Grand}, F.~A. {G{\'o}mez}, F.~{Marinacci}, R.~{Pakmor},
  V.~{Springel}, D.~J.~R. {Campbell} et~al., \emph{{The Auriga Project: the
  properties and formation mechanisms of disc galaxies across cosmic time}},
  \href{https://doi.org/10.1093/mnras/stx071}{\emph{\mnras} {\bfseries 467}
  (2017) 179} [\href{https://arxiv.org/abs/1610.01159}{{\ttfamily
  1610.01159}}].

\bibitem{Nunez-Castineyra2023}
A.~{Nu{\~n}ez-Casti{\~n}eyra}, E.~{Nezri}, P.~{Mollitor}, J.~{Devriendt} and
  R.~{Teyssier}, \emph{{Cosmological simulations of the same spiral galaxy:
  connecting the dark matter distribution of the host halo with the subgrid
  baryonic physics}},
  \href{https://doi.org/10.1088/1475-7516/2023/05/012}{\emph{\jcap} {\bfseries
  2023} (2023) 012} [\href{https://arxiv.org/abs/2301.06189}{{\ttfamily
  2301.06189}}].

\bibitem{Wetzel2016}
A.~R. {Wetzel}, P.~F. {Hopkins}, J.-h. {Kim}, C.-A. {Faucher-Gigu{\`e}re},
  D.~{Kere{\v{s}}} and E.~{Quataert}, \emph{{Reconciling Dwarf Galaxies with
  {\ensuremath{\Lambda}}CDM Cosmology: Simulating a Realistic Population of
  Satellites around a Milky Way-mass Galaxy}},
  \href{https://doi.org/10.3847/2041-8205/827/2/L23}{\emph{\apjl} {\bfseries
  827} (2016) L23} [\href{https://arxiv.org/abs/1602.05957}{{\ttfamily
  1602.05957}}].

\bibitem{Teyssier:2001cp}
R.~Teyssier, \emph{{Cosmological hydrodynamics with adaptive mesh refinement: a
  new high resolution code called ramses}},
  \href{https://doi.org/10.1051/0004-6361:20011817}{\emph{Astron. Astrophys.}
  {\bfseries 385} (2002) 337}
  [\href{https://arxiv.org/abs/astro-ph/0111367}{{\ttfamily
  astro-ph/0111367}}].

\bibitem{HahnAbel2013}
O.~{Hahn} and T.~{Abel}, ``{MUSIC: MUlti-Scale Initial Conditions}.''
  Astrophysics Source Code Library, record ascl:1311.011, Nov., 2013.

\bibitem{Mollitor:2014ara}
P.~Mollitor, E.~Nezri and R.~Teyssier, \emph{{Baryonic and dark matter
  distribution in cosmological simulations of spiral galaxies}},
  \href{https://doi.org/10.1093/mnras/stu2466}{\emph{MNRAS} {\bfseries 447}
  (2015) 1353} [\href{https://arxiv.org/abs/1405.4318}{{\ttfamily 1405.4318}}].

\bibitem{Marinacci:2013mha}
F.~Marinacci, R.~Pakmor and V.~Springel, \emph{{The formation of disc galaxies
  in high resolution moving-mesh cosmological simulations}},
  \href{https://doi.org/10.1093/mnras/stt2003}{\emph{MNRAS} {\bfseries 437}
  (2014) 1750} [\href{https://arxiv.org/abs/1305.5360}{{\ttfamily 1305.5360}}].

\bibitem{Kennicut1998}
J.~{Kennicutt}, Robert~C., \emph{{The Global Schmidt Law in Star-forming
  Galaxies}}, \href{https://doi.org/10.1086/305588}{\emph{\apj} {\bfseries 498}
  (1998) 541} [\href{https://arxiv.org/abs/astro-ph/9712213}{{\ttfamily
  astro-ph/9712213}}].

\bibitem{Teyssier2013}
R.~{Teyssier}, A.~{Pontzen}, Y.~{Dubois} and J.~I. {Read}, \emph{{Cusp-core
  transformations in dwarf galaxies: observational predictions}},
  \href{https://doi.org/10.1093/mnras/sts563}{\emph{\mnras} {\bfseries 429}
  (2013) 3068} [\href{https://arxiv.org/abs/1206.4895}{{\ttfamily 1206.4895}}].

\bibitem{Federrath2012}
C.~{Federrath} and R.~S. {Klessen}, \emph{{The Star Formation Rate of Turbulent
  Magnetized Clouds: Comparing Theory, Simulations, and Observations}},
  \href{https://doi.org/10.1088/0004-637X/761/2/156}{\emph{\apj} {\bfseries
  761} (2012) 156} [\href{https://arxiv.org/abs/1209.2856}{{\ttfamily
  1209.2856}}].

\bibitem{Krumholz2005}
M.~R. {Krumholz} and C.~F. {McKee}, \emph{{A General Theory of
  Turbulence-regulated Star Formation, from Spirals to Ultraluminous Infrared
  Galaxies}}, \href{https://doi.org/10.1086/431734}{\emph{\apj} {\bfseries 630}
  (2005) 250} [\href{https://arxiv.org/abs/astro-ph/0505177}{{\ttfamily
  astro-ph/0505177}}].

\bibitem{Kimm2015}
T.~{Kimm}, R.~{Cen}, J.~{Devriendt}, Y.~{Dubois} and A.~{Slyz}, \emph{{Towards
  simulating star formation in turbulent high-z galaxies with mechanical
  supernova feedback}},
  \href{https://doi.org/10.1093/mnras/stv1211}{\emph{\mnras} {\bfseries 451}
  (2015) 2900} [\href{https://arxiv.org/abs/1501.05655}{{\ttfamily
  1501.05655}}].

\bibitem{SKIRTCamps2015}
P.~{Camps} and M.~{Baes}, \emph{{SKIRT: An advanced dust radiative transfer
  code with a user-friendly architecture}},
  \href{https://doi.org/10.1016/j.ascom.2014.10.004}{\emph{Astronomy and
  Computing} {\bfseries 9} (2015) 20}
  [\href{https://arxiv.org/abs/1410.1629}{{\ttfamily 1410.1629}}].

\bibitem{Springel:2008cc}
V.~Springel, J.~Wang, M.~Vogelsberger, A.~Ludlow, A.~Jenkins, A.~Helmi et~al.,
  \emph{{The Aquarius Project: the subhalos of galactic halos}},
  \href{https://doi.org/10.1111/j.1365-2966.2008.14066.x}{\emph{Mon. Not. Roy.
  Astron. Soc.} {\bfseries 391} (2008) 1685}
  [\href{https://arxiv.org/abs/0809.0898}{{\ttfamily 0809.0898}}].

\bibitem{binney2011galactic}
J.~Binney and S.~Tremaine, \emph{Galactic dynamics}, vol.~13. Princeton
  university press, 2011.

\bibitem{HuchraGeller1982}
J.~P. {Huchra} and M.~J. {Geller}, \emph{{Groups of Galaxies. I. Nearby
  groups}}, \href{https://doi.org/10.1086/160000}{\emph{\apj} {\bfseries 257}
  (1982) 423}.

\bibitem{Davis1985}
M.~{Davis}, G.~{Efstathiou}, C.~S. {Frenk} and S.~D.~M. {White}, \emph{{The
  evolution of large-scale structure in a universe dominated by cold dark
  matter}}, \href{https://doi.org/10.1086/163168}{\emph{\apj} {\bfseries 292}
  (1985) 371}.

\bibitem{Maciejewski2009}
M.~{Maciejewski}, S.~{Colombi}, V.~{Springel}, C.~{Alard} and F.~R. {Bouchet},
  \emph{{Phase-space structures - II. Hierarchical Structure Finder}},
  \href{https://doi.org/10.1111/j.1365-2966.2009.14825.x}{\emph{\mnras}
  {\bfseries 396} (2009) 1329}
  [\href{https://arxiv.org/abs/0812.0288}{{\ttfamily 0812.0288}}].

\bibitem{Behroozi2012Rockstar}
P.~{Behroozi}, R.~{Wechsler} and H.-Y. {Wu}, ``{Rockstar: Phase-space halo
  finder}.'' Astrophysics Source Code Library, record ascl:1210.008, Oct.,
  2012.

\bibitem{Vladimir2025}
Z.~{Vladimir}, C.~{Osinga}, B.~{Diemer}, E.~M. {Salazar} and E.~{Rozo},
  \emph{{Distinguishing Orbiting and Infalling Dark Matter Particles with
  Machine Learning}},
  \href{https://doi.org/10.48550/arXiv.2506.09146}{\emph{arXiv e-prints} (2025)
  arXiv:2506.09146} [\href{https://arxiv.org/abs/2506.09146}{{\ttfamily
  2506.09146}}].

\bibitem{Springel2001SUBFIND}
V.~{Springel}, S.~D.~M. {White}, G.~{Tormen} and G.~{Kauffmann},
  \emph{{Populating a cluster of galaxies - I. Results at z=0}},
  \href{https://doi.org/10.1046/j.1365-8711.2001.04912.x}{\emph{\mnras}
  {\bfseries 328} (2001) 726}
  [\href{https://arxiv.org/abs/astro-ph/0012055}{{\ttfamily
  astro-ph/0012055}}].

\bibitem{Gao2011}
L.~{Gao}, C.~S. {Frenk}, M.~{Boylan-Kolchin}, A.~{Jenkins}, V.~{Springel} and
  S.~D.~M. {White}, \emph{{The statistics of the subhalo abundance of dark
  matter haloes}},
  \href{https://doi.org/10.1111/j.1365-2966.2010.17601.x}{\emph{\mnras}
  {\bfseries 410} (2011) 2309}
  [\href{https://arxiv.org/abs/1006.2882}{{\ttfamily 1006.2882}}].

\bibitem{Errani2018}
R.~{Errani}, J.~{Pe{\~n}arrubia} and M.~G. {Walker}, \emph{{Systematics in
  virial mass estimators for pressure-supported systems}},
  \href{https://doi.org/10.1093/mnras/sty2505}{\emph{\mnras} {\bfseries 481}
  (2018) 5073} [\href{https://arxiv.org/abs/1805.00484}{{\ttfamily
  1805.00484}}].

\bibitem{Nadler2020}
E.~O. {Nadler}, R.~H. {Wechsler}, K.~{Bechtol}, Y.~Y. {Mao}, G.~{Green},
  A.~{Drlica-Wagner} et~al., \emph{{Milky Way Satellite Census. II. Galaxy-Halo
  Connection Constraints Including the Impact of the Large Magellanic Cloud}},
  \href{https://doi.org/10.3847/1538-4357/ab846a}{\emph{\apj} {\bfseries 893}
  (2020) 48} [\href{https://arxiv.org/abs/1912.03303}{{\ttfamily 1912.03303}}].

\bibitem{Behroozi2013}
P.~S. {Behroozi}, R.~H. {Wechsler} and C.~{Conroy}, \emph{{The Average Star
  Formation Histories of Galaxies in Dark Matter Halos from z = 0-8}},
  \href{https://doi.org/10.1088/0004-637X/770/1/57}{\emph{\apj} {\bfseries 770}
  (2013) 57} [\href{https://arxiv.org/abs/1207.6105}{{\ttfamily 1207.6105}}].

\bibitem{Moster2010}
B.~P. {Moster}, R.~S. {Somerville}, C.~{Maulbetsch}, F.~C. {van den Bosch},
  A.~V. {Macci{\`o}}, T.~{Naab} et~al., \emph{{Constraints on the Relationship
  between Stellar Mass and Halo Mass at Low and High Redshift}},
  \href{https://doi.org/10.1088/0004-637X/710/2/903}{\emph{\apj} {\bfseries
  710} (2010) 903} [\href{https://arxiv.org/abs/0903.4682}{{\ttfamily
  0903.4682}}].

\bibitem{Danieli2023}
S.~{Danieli}, J.~E. {Greene}, S.~{Carlsten}, F.~{Jiang}, R.~{Beaton} and A.~D.
  {Goulding}, \emph{{ELVES. IV. The Satellite Stellar-to-halo Mass Relation
  Beyond the Milky Way}},
  \href{https://doi.org/10.3847/1538-4357/acefbd}{\emph{\apj} {\bfseries 956}
  (2023) 6} [\href{https://arxiv.org/abs/2210.14233}{{\ttfamily 2210.14233}}].

\bibitem{RodriguezPuebla2012}
A.~{Rodr{\'\i}guez-Puebla}, N.~{Drory} and V.~{Avila-Reese}, \emph{{The
  Stellar-Subhalo Mass Relation of Satellite Galaxies}},
  \href{https://doi.org/10.1088/0004-637X/756/1/2}{\emph{\apj} {\bfseries 756}
  (2012) 2} [\href{https://arxiv.org/abs/1204.0804}{{\ttfamily 1204.0804}}].

\bibitem{Applebaum2021}
E.~{Applebaum}, A.~M. {Brooks}, C.~R. {Christensen}, F.~{Munshi}, T.~R.
  {Quinn}, S.~{Shen} et~al., \emph{{Ultrafaint Dwarfs in a Milky Way Context:
  Introducing the Mint Condition DC Justice League Simulations}},
  \href{https://doi.org/10.3847/1538-4357/abcafa}{\emph{\apj} {\bfseries 906}
  (2021) 96} [\href{https://arxiv.org/abs/2008.11207}{{\ttfamily 2008.11207}}].

\bibitem{Diemand:2007qr}
J.~Diemand, M.~Kuhlen and P.~Madau, \emph{{Formation and evolution of galaxy
  dark matter halos and their substructure}},
  \href{https://doi.org/10.1086/520573}{\emph{Astrophys. J.} {\bfseries 667}
  (2007) 859} [\href{https://arxiv.org/abs/astro-ph/0703337}{{\ttfamily
  astro-ph/0703337}}].

\bibitem{Wang2020}
J.~{Wang}, S.~{Bose}, C.~S. {Frenk}, L.~{Gao}, A.~{Jenkins}, V.~{Springel}
  et~al., \emph{{Universal structure of dark matter haloes over a mass range of
  20 orders of magnitude}},
  \href{https://doi.org/10.1038/s41586-020-2642-9}{\emph{\nat} {\bfseries 585}
  (2020) 39} [\href{https://arxiv.org/abs/1911.09720}{{\ttfamily 1911.09720}}].

\bibitem{PressSchechter1974}
W.~H. {Press} and P.~{Schechter}, \emph{{Formation of Galaxies and Clusters of
  Galaxies by Self-Similar Gravitational Condensation}},
  \href{https://doi.org/10.1086/152650}{\emph{\apj} {\bfseries 187} (1974)
  425}.

\bibitem{Bond1991}
J.~R. {Bond}, S.~{Cole}, G.~{Efstathiou} and N.~{Kaiser}, \emph{{Excursion Set
  Mass Functions for Hierarchical Gaussian Fluctuations}},
  \href{https://doi.org/10.1086/170520}{\emph{\apj} {\bfseries 379} (1991)
  440}.

\bibitem{LaceyCole1993}
C.~{Lacey} and S.~{Cole}, \emph{{Merger rates in hierarchical models of galaxy
  formation}}, \href{https://doi.org/10.1093/mnras/262.3.627}{\emph{\mnras}
  {\bfseries 262} (1993) 627}.

\bibitem{Tormen1998}
G.~{Tormen}, A.~{Diaferio} and D.~{Syer}, \emph{{Survival of substructure
  within dark matter haloes}},
  \href{https://doi.org/10.1046/j.1365-8711.1998.01775.x}{\emph{\mnras}
  {\bfseries 299} (1998) 728}
  [\href{https://arxiv.org/abs/astro-ph/9712222}{{\ttfamily
  astro-ph/9712222}}].

\bibitem{Zentner2007}
A.~R. {Zentner}, \emph{{The Excursion Set Theory of Halo Mass Functions, Halo
  Clustering, and Halo Growth}},
  \href{https://doi.org/10.1142/S0218271807010511}{\emph{International Journal
  of Modern Physics D} {\bfseries 16} (2007) 763}
  [\href{https://arxiv.org/abs/astro-ph/0611454}{{\ttfamily
  astro-ph/0611454}}].

\bibitem{SomervillePrimack1999}
R.~S. {Somerville} and J.~R. {Primack}, \emph{{Semi-analytic modelling of
  galaxy formation: the local Universe}},
  \href{https://doi.org/10.1046/j.1365-8711.1999.03032.x}{\emph{\mnras}
  {\bfseries 310} (1999) 1087}
  [\href{https://arxiv.org/abs/astro-ph/9802268}{{\ttfamily
  astro-ph/9802268}}].

\bibitem{Jeon2025}
S.~{Jeon}, S.~K. {Yi}, E.~{Contini}, Y.~{Dubois}, S.~{Han}, K.~{Kraljic}
  et~al., \emph{{Born to be Starless: Revisiting the Missing Satellite
  Problem}}, \href{https://doi.org/10.48550/arXiv.2506.09152}{\emph{arXiv
  e-prints} (2025) arXiv:2506.09152}
  [\href{https://arxiv.org/abs/2506.09152}{{\ttfamily 2506.09152}}].

\bibitem{Governato:2009bg}
F.~Governato et~al., \emph{{At the heart of the matter: the origin of bulgeless
  dwarf galaxies and Dark Matter cores}},
  \href{https://doi.org/10.1038/nature08640}{\emph{Nature} {\bfseries 463}
  (2010) 203} [\href{https://arxiv.org/abs/0911.2237}{{\ttfamily 0911.2237}}].

\bibitem{Pontzen2012}
A.~{Pontzen} and F.~{Governato}, \emph{{How supernova feedback turns dark
  matter cusps into cores}},
  \href{https://doi.org/10.1111/j.1365-2966.2012.20571.x}{\emph{\mnras}
  {\bfseries 421} (2012) 3464}
  [\href{https://arxiv.org/abs/1106.0499}{{\ttfamily 1106.0499}}].

\bibitem{El-Zan2016}
A.~A. {El-Zant}, J.~{Freundlich} and F.~{Combes}, \emph{{From cusps to cores: a
  stochastic model}},
  \href{https://doi.org/10.1093/mnras/stw1398}{\emph{\mnras} {\bfseries 461}
  (2016) 1745} [\href{https://arxiv.org/abs/1603.00526}{{\ttfamily
  1603.00526}}].

\bibitem{Freundlich2020}
J.~{Freundlich}, A.~{Dekel}, F.~{Jiang}, G.~{Ishai}, N.~{Cornuault},
  S.~{Lapiner} et~al., \emph{{A model for core formation in dark matter haloes
  and ultra-diffuse galaxies by outflow episodes}},
  \href{https://doi.org/10.1093/mnras/stz3306}{\emph{\mnras} {\bfseries 491}
  (2020) 4523} [\href{https://arxiv.org/abs/1907.11726}{{\ttfamily
  1907.11726}}].

\bibitem{Li2023}
Z.~{Li}, A.~{Dekel}, N.~{Mandelker}, J.~{Freundlich} and T.~L.
  {Fran{\c{c}}ois}, \emph{{The response of dark matter haloes to gas ejection:
  CuspCore II}}, \href{https://doi.org/10.1093/mnras/stac3233}{\emph{\mnras}
  {\bfseries 518} (2023) 5356}
  [\href{https://arxiv.org/abs/2206.07069}{{\ttfamily 2206.07069}}].

\bibitem{Tollet2017}
{\'E}.~{Tollet}, A.~{Cattaneo}, G.~A. {Mamon}, T.~{Moutard} and F.~C. {van den
  Bosch}, \emph{{On stellar mass loss from galaxies in groups and clusters}},
  \href{https://doi.org/10.1093/mnras/stx1840}{\emph{\mnras} {\bfseries 471}
  (2017) 4170} [\href{https://arxiv.org/abs/1707.06264}{{\ttfamily
  1707.06264}}].

\bibitem{DiCintio2017}
A.~{Di Cintio}, C.~B. {Brook}, A.~A. {Dutton}, A.~V. {Macci{\`o}}, A.~{Obreja}
  and A.~{Dekel}, \emph{{NIHAO - XI. Formation of ultra-diffuse galaxies by
  outflows}}, \href{https://doi.org/10.1093/mnrasl/slw210}{\emph{\mnras}
  {\bfseries 466} (2017) L1}
  [\href{https://arxiv.org/abs/1608.01327}{{\ttfamily 1608.01327}}].

\bibitem{Fitts2017}
A.~{Fitts}, M.~{Boylan-Kolchin}, O.~D. {Elbert}, J.~S. {Bullock}, P.~F.
  {Hopkins}, J.~{O{\~n}orbe} et~al., \emph{{fire in the field: simulating the
  threshold of galaxy formation}},
  \href{https://doi.org/10.1093/mnras/stx1757}{\emph{\mnras} {\bfseries 471}
  (2017) 3547} [\href{https://arxiv.org/abs/1611.02281}{{\ttfamily
  1611.02281}}].

\bibitem{Dekel2021}
A.~{Dekel}, J.~{Freundlich}, F.~{Jiang}, S.~{Lapiner}, A.~{Burkert},
  D.~{Ceverino} et~al., \emph{{Core formation in high-z massive haloes: heating
  by post-compaction satellites and response to AGN outflows}},
  \href{https://doi.org/10.1093/mnras/stab2416}{\emph{\mnras} {\bfseries 508}
  (2021) 999} [\href{https://arxiv.org/abs/2106.01378}{{\ttfamily
  2106.01378}}].

\bibitem{Hashim2023}
M.~{Hashim}, A.~A. {El-Zant}, J.~{Freundlich}, J.~I. {Read} and F.~{Combes},
  \emph{{Halo heating from fluctuating gas in a model dwarf}},
  \href{https://doi.org/10.1093/mnras/stad581}{\emph{\mnras} {\bfseries 521}
  (2023) 772} [\href{https://arxiv.org/abs/2209.08631}{{\ttfamily
  2209.08631}}].

\bibitem{Weisz2014}
D.~R. {Weisz}, A.~E. {Dolphin}, E.~D. {Skillman}, J.~{Holtzman}, K.~M.
  {Gilbert}, J.~J. {Dalcanton} et~al., \emph{{The Star Formation Histories of
  Local Group Dwarf Galaxies. I. Hubble Space Telescope/Wide Field Planetary
  Camera 2 Observations}},
  \href{https://doi.org/10.1088/0004-637X/789/2/147}{\emph{\apj} {\bfseries
  789} (2014) 147} [\href{https://arxiv.org/abs/1404.7144}{{\ttfamily
  1404.7144}}].

\bibitem{Jiang2019}
F.~{Jiang}, A.~{Dekel}, J.~{Freundlich}, A.~J. {Romanowsky}, A.~A. {Dutton},
  A.~V. {Macci{\`o}} et~al., \emph{{Formation of ultra-diffuse galaxies in the
  field and in galaxy groups}},
  \href{https://doi.org/10.1093/mnras/stz1499}{\emph{\mnras} {\bfseries 487}
  (2019) 5272} [\href{https://arxiv.org/abs/1811.10607}{{\ttfamily
  1811.10607}}].

\bibitem{ErraniPenarrubia2020}
R.~{Errani} and J.~{Pe{\~n}arrubia}, \emph{{Can tides disrupt cold dark matter
  subhaloes?}}, \href{https://doi.org/10.1093/mnras/stz3349}{\emph{\mnras}
  {\bfseries 491} (2020) 4591}
  [\href{https://arxiv.org/abs/1906.01642}{{\ttfamily 1906.01642}}].

\bibitem{Read2019}
J.~I. {Read}, M.~G. {Walker} and P.~{Steger}, \emph{{Dark matter heats up in
  dwarf galaxies}}, \href{https://doi.org/10.1093/mnras/sty3404}{\emph{\mnras}
  {\bfseries 484} (2019) 1401}
  [\href{https://arxiv.org/abs/1808.06634}{{\ttfamily 1808.06634}}].

\bibitem{Oh2015}
S.-H. {Oh}, D.~A. {Hunter}, E.~{Brinks}, B.~G. {Elmegreen}, A.~{Schruba},
  F.~{Walter} et~al., \emph{{High-resolution Mass Models of Dwarf Galaxies from
  LITTLE THINGS}},
  \href{https://doi.org/10.1088/0004-6256/149/6/180}{\emph{\aj} {\bfseries 149}
  (2015) 180} [\href{https://arxiv.org/abs/1502.01281}{{\ttfamily
  1502.01281}}].

\bibitem{Vitral2024}
E.~{Vitral}, R.~P. {van der Marel}, S.~T. {Sohn}, M.~{Libralato}, A.~{del
  Pino}, L.~L. {Watkins} et~al., \emph{{HSTPROMO Internal Proper-motion
  Kinematics of Dwarf Spheroidal Galaxies. I. Velocity Anisotropy and Dark
  Matter Cusp Slope of Draco}},
  \href{https://doi.org/10.3847/1538-4357/ad571c}{\emph{\apj} {\bfseries 970}
  (2024) 1} [\href{https://arxiv.org/abs/2407.07769}{{\ttfamily 2407.07769}}].

\bibitem{Pascale2025}
R.~{Pascale}, C.~{Nipoti}, F.~{Calura} and A.~{Della Croce}, \emph{{Leo I: the
  classical dwarf spheroidal galaxy with the highest dark-matter density}},
  \href{https://doi.org/10.48550/arXiv.2506.13847}{\emph{arXiv e-prints} (2025)
  arXiv:2506.13847} [\href{https://arxiv.org/abs/2506.13847}{{\ttfamily
  2506.13847}}].

\bibitem{Yang2025}
H.~{Yang}, W.~{Wang}, L.~{Zhu}, T.~S. {Li}, S.~E. {Koposov}, J.~{Han} et~al.,
  \emph{{The dark matter content of Milky Way dwarf spheroidal galaxies: Draco,
  Sextans and Ursa Minor}},
  \href{https://doi.org/10.48550/arXiv.2507.02284}{\emph{arXiv e-prints} (2025)
  arXiv:2507.02284} [\href{https://arxiv.org/abs/2507.02284}{{\ttfamily
  2507.02284}}].

\bibitem{Dutton:2018nop}
A.~A. Dutton, A.~V. Macciò, T.~Buck, K.~L. Dixon, M.~Blank and A.~Obreja,
  \emph{{NIHAO XX: the impact of the star formation threshold on the
  cusp–core transformation of cold dark matter haloes}},
  \href{https://doi.org/10.1093/mnras/stz889}{\emph{MNRAS} {\bfseries 486}
  (2019) 655} [\href{https://arxiv.org/abs/1811.10625}{{\ttfamily
  1811.10625}}].

\bibitem{Orkey2021}
M.~D.~A. {Orkney}, J.~I. {Read}, M.~P. {Rey}, I.~{Nasim}, A.~{Pontzen},
  O.~{Agertz} et~al., \emph{{EDGE: two routes to dark matter core formation in
  ultra-faint dwarfs}},
  \href{https://doi.org/10.1093/mnras/stab1066}{\emph{\mnras} {\bfseries 504}
  (2021) 3509} [\href{https://arxiv.org/abs/2101.02688}{{\ttfamily
  2101.02688}}].

\bibitem{Muni2025}
C.~{Muni}, A.~{Pontzen}, J.~I. {Read}, O.~{Agertz}, M.~P. {Rey}, E.~{Taylor}
  et~al., \emph{{EDGE: dark matter core creation depends on the timing of star
  formation}}, \href{https://doi.org/10.1093/mnras/stae2748}{\emph{\mnras}
  {\bfseries 536} (2025) 314}
  [\href{https://arxiv.org/abs/2407.14579}{{\ttfamily 2407.14579}}].

\bibitem{Borukhovetskaya2022}
A.~{Borukhovetskaya}, J.~F. {Navarro}, R.~{Errani} and A.~{Fattahi},
  \emph{{Galactic tides and the Crater II dwarf spheroidal: a challenge to
  LCDM?}}, \href{https://doi.org/10.1093/mnras/stac653}{\emph{\mnras}
  {\bfseries 512} (2022) 5247}
  [\href{https://arxiv.org/abs/2112.01540}{{\ttfamily 2112.01540}}].

\bibitem{Dehnen2002}
W.~{Dehnen}, \emph{{A Hierarchical <E10>O</E10>(N) Force Calculation
  Algorithm}}, \href{https://doi.org/10.1006/jcph.2002.7026}{\emph{Journal of
  Computational Physics} {\bfseries 179} (2002) 27}
  [\href{https://arxiv.org/abs/astro-ph/0202512}{{\ttfamily
  astro-ph/0202512}}].

\end{thebibliography}\endgroup

\bibliographystyle{JHEP}
\appendix

\begin{figure}[t]
\centering
\begin{subfigure}[b]{0.45\linewidth}
\includegraphics[width=\linewidth]{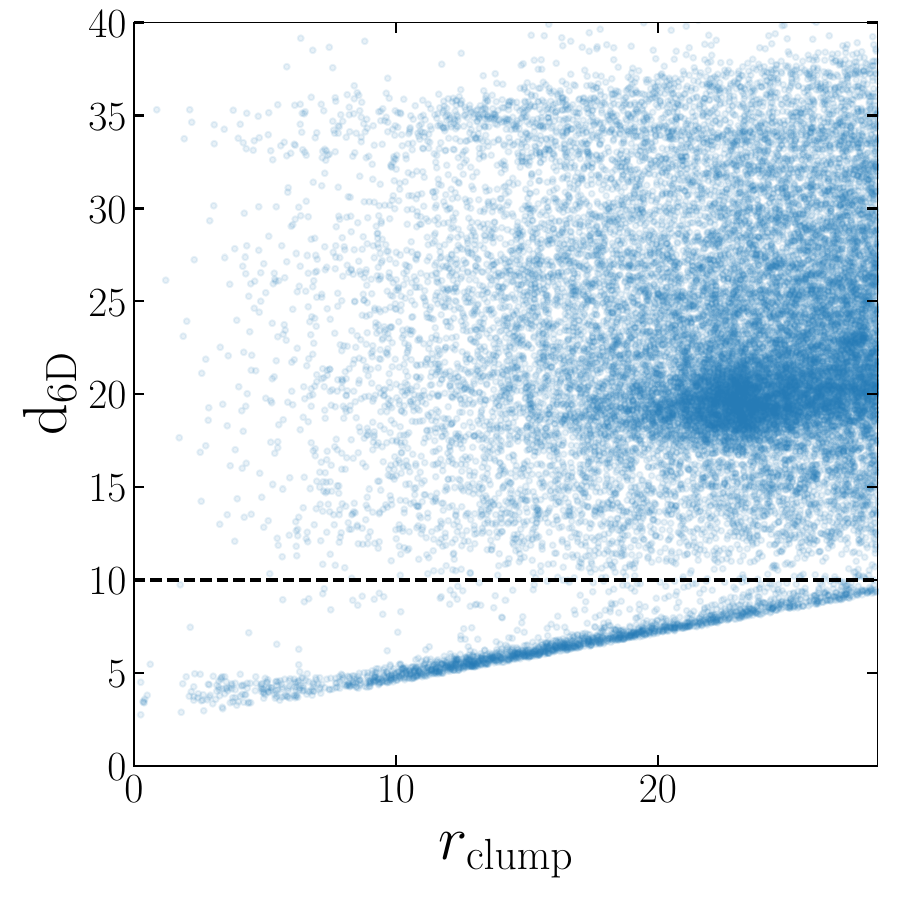}
\end{subfigure}
\begin{subfigure}[b]{0.45\linewidth}
\includegraphics[width=\linewidth]{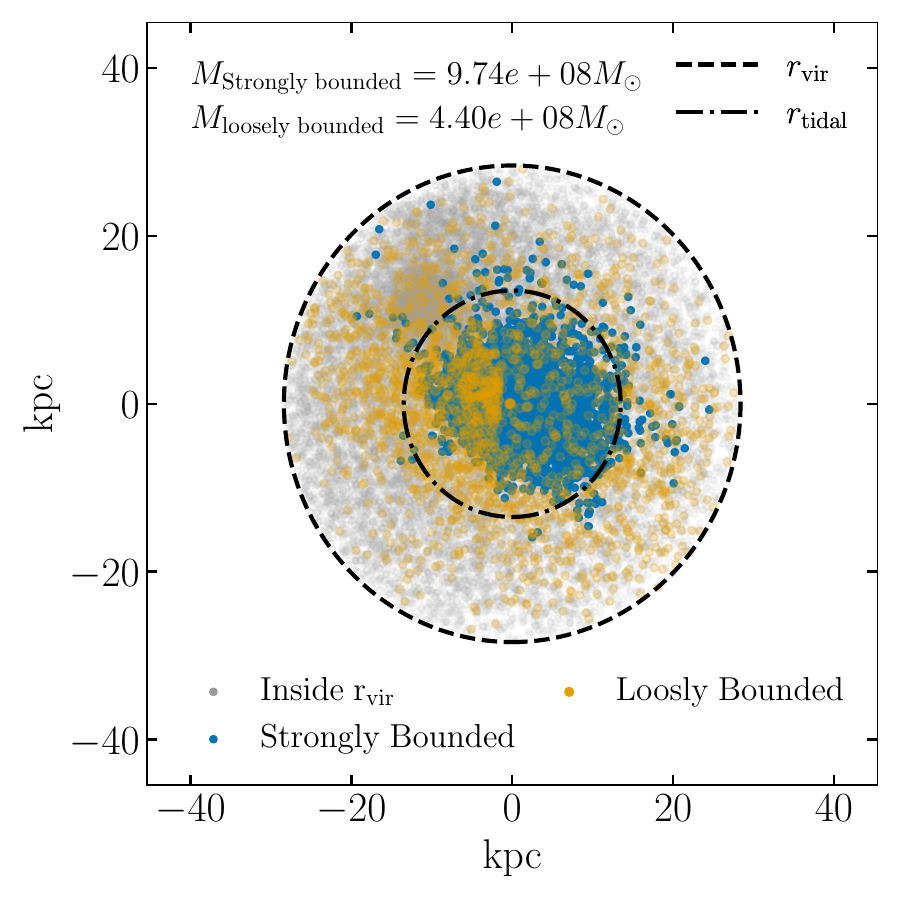}
\end{subfigure}
\caption{Example of the six-dimensional phase-space distance used to classify loosely bound particles for a single clump. The left panel shows the $d_{6D}$ values of particles located within the virial radius that were not classified as strongly bound. The dashed line marks the empirically chosen threshold below which particles are identified as loosely bound. The right panel shows the spatial distribution of the four particle classes: strongly bound (blue), loosely bound (orange), and the rest of the particles inside $r_{\rm vir}$ (gray). The positions are projected onto the plane defined by the velocity and position vectors of the clump. The dashed circle indicates the virial radius, the dash-dotted line the tidal radius, and the dotted line the shock radius.}
\label{fig:clumpexample}%
\end{figure}
\section{Clump particle identification}\label{app:particleselection}

For each selected clump, the analysis was conducted in the frame of reference of the subhalo itself, with the local centre-of-mass velocity subtracted to isolate internal dynamics. Rather than relying on the global simulation potential, a corrected local gravitational potential was reconstructed\footnote{The gravitational potential was computed using the \texttt{unsio} library, employing the \texttt{falcON} method developed by Walter Dehnen \cite{Dehnen2002}. This method uses a fast multipole expansion to efficiently estimate the gravitational potential of the system.}. For each subhalo, the radial profile of the host halo background potential was computed and subtracted from the total potential. This procedure isolates the contribution of the subhalo, mitigating contamination from the strong tidal forces and asymmetric distortions induced by the host halo. Consequently, the method enables a more faithful identification of the true gravitational well of the subhalo, even in regions where gravitational harassment and environmental asymmetries are significant. This refinement permits the gravitational potential well of each structure to be isolated with far greater fidelity than standard spherically symmetric methods.

The particles were classified as strongly bound based on two physical criteria. First, particles were required to have negative total energy, evaluated relative to the corrected local gravitational potential. Second, the directionality of their motion and dynamical response was considered: particles were selected if the angle between their gravitational acceleration and the direction toward the centre of the subhalo was smaller than $90^\circ$, indicating that they were being pulled inward toward the subhalo centre, or, if this condition was not satisfied, if the angle between their velocity vector and the centre direction was smaller than $135^\circ$ while still meeting the energy condition, indicating that they were moving outward but still sufficiently oriented toward the subhalo to remain gravitationally influenced. This additional constraint was introduced to prevent the erroneous classification of particles that were already escaping and unlikely to be pulled back in as bound material.
Next, to classify the loosely bound particles in the outskirts of clumps that follow trajectories coherent with the motion of the clump, an additional phase‑space selection is applied by computing a dimensionless six‑dimensional distance each nearby particle not previously selected as :

\begin{equation}
d_{6D,i} = \sqrt{ \frac{|\mathbf{r}_i - \mathbf{r}_{\mathrm{bounded}}|^2}{\sigma_x^2} + \frac{|\mathbf{v}_i - \mathbf{v}_{\mathrm{bounded}}|^2}{\sigma_v^2} }.
\end{equation}

Here, $d_{6D,i}$ is calculated for the particle $i$; $\mathbf{r}_i$ and $\mathbf{v}_i$ are its position and velocity vectors; $\mathbf{r}_{\mathrm{bounded}}$ and $\mathbf{v}_{\mathrm{bounded}}$ are the mean position and velocity of the group of particles previously identified as strongly bound; and $\sigma_x$ and $\sigma_v$ are the spatial and velocity dispersion of that group, respectively. Particles with $d_{6D} < d_{\mathrm{thresh}}$ (with $d_{\mathrm{thresh}} = 10$, empirically chosen to select a group that separates from the rest of the nearby particles in the $d_{6D,i}$–$r_{\rm clump}$ plane) are identified as part of the loosely bound tail. The application of this phase‑space tail selection improves the completeness of the clump sample, particularly in cases where tidal stripping produces asymmetric outskirts that are not well captured by energy and local density criteria alone.
Figure \ref{fig:clumpexample} shows an example of the spatial distribution of the different particle classes, projected onto the plane defined by the velocity and position vectors of a selected clump. Typically, the tidal radius encloses a smaller volume containing primarily bound particles. In contrast, the virial radius, as reported by the halo finder, defines a much larger region. When used purely as a geometric boundary, it tends to include a significant number of background particles that are not dynamically associated with the clump.


\section{Examples of dark matter profiles}
\begin{figure}[t]
\centering
\begin{subfigure}[b]{0.45\linewidth}
\includegraphics[width=\linewidth]{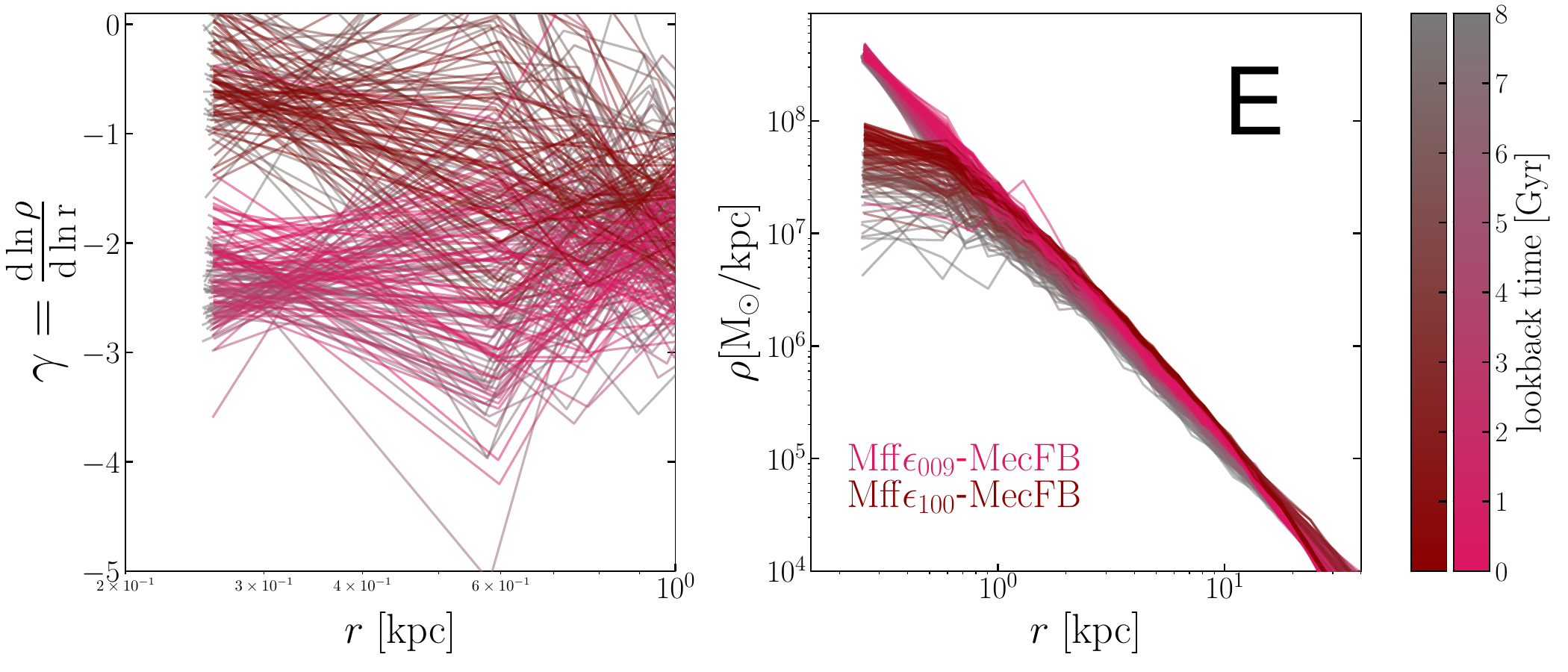}
\end{subfigure}
\begin{subfigure}[b]{0.45\linewidth}
\includegraphics[width=\linewidth]{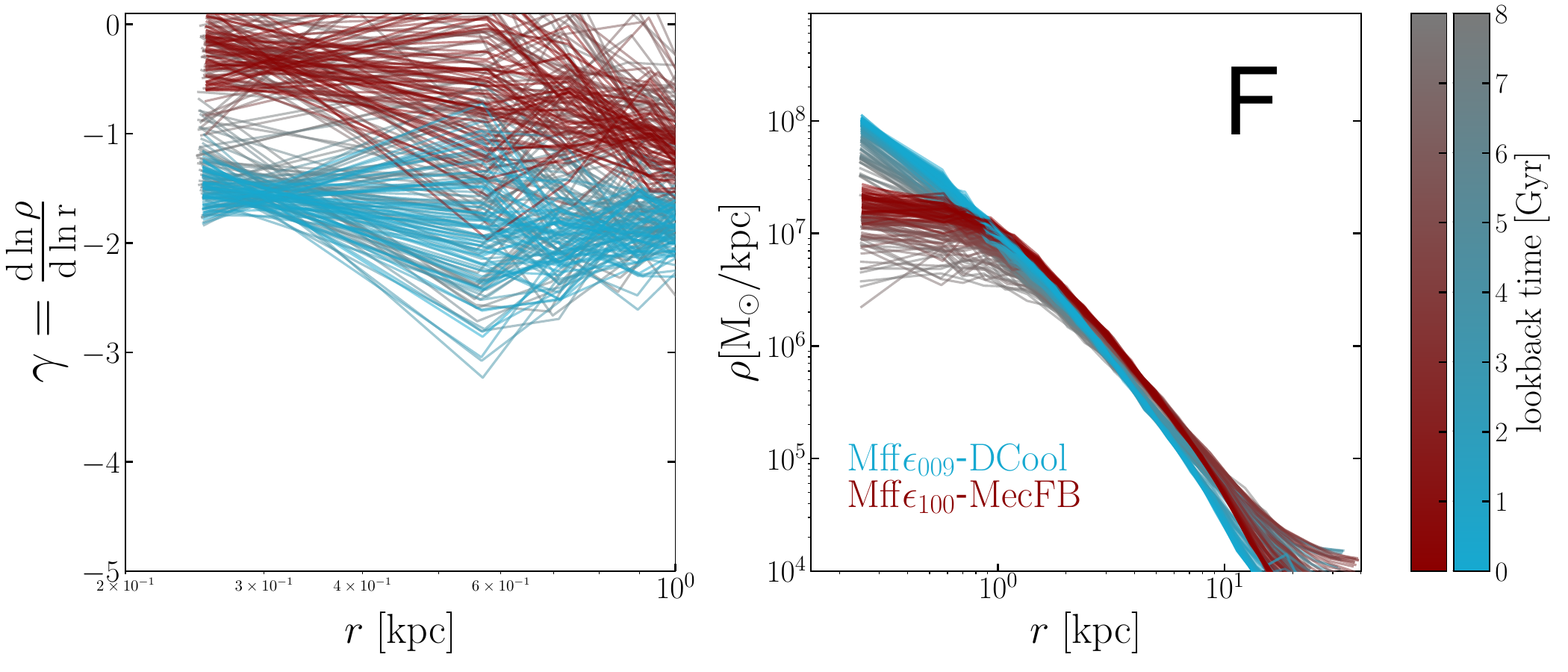}
\end{subfigure}
\caption{Dark matter density profiles of halos E and F, showing the two realizations with the most extreme inner slopes ($\gamma$). Left panels: radial slope $\gamma = d\ln \rho / d\ln r$ within $r < 1$ kpc. Right panels: full density profiles. Colors indicate simulation and time.}
\label{fig:exampleprofiles}%
\end{figure}

As a complement to Figure \ref{fig:evo}, the dark matter profiles of the two inner-slope extremes of each satellite halo (E and F) are shown over time in Figure \ref{fig:exampleprofiles}. The profiles are color-coded according to simulation and time, as indicated by the color bar. The earliest profile corresponds to 8 Gyr ago, which marks the starting point of the analysis presented above.

For each simulation, the left panel shows the numerical derivative of the natural logarithm of the density with respect to the natural logarithm of the radius, i.e. the radial slope $\gamma = d\ln \rho / d\ln r$ of a power law. This quantity is shown only within $r < 1$~kpc. The right panel for each subhalo shows the full density profile for the two realizations corresponding to the extreme values of $\gamma$.

These profiles reveal that the two cases exhibiting cores correspond to the same run, \rouge, whereas the cuspy profiles are found in different runs: \rose\ for halo E and \blue\ for halo F. A more notable feature is the larger dispersion in the central density of the cored profiles compared to the cuspy ones, indicating that cores are less stable than cusps in satellite galaxies (see Section \ref{sec:Discussion}).

\section{Orbits of example subhalos.}

\begin{figure}[t]
\centering
\begin{subfigure}[b]{0.75\linewidth}
\includegraphics[width=\linewidth]{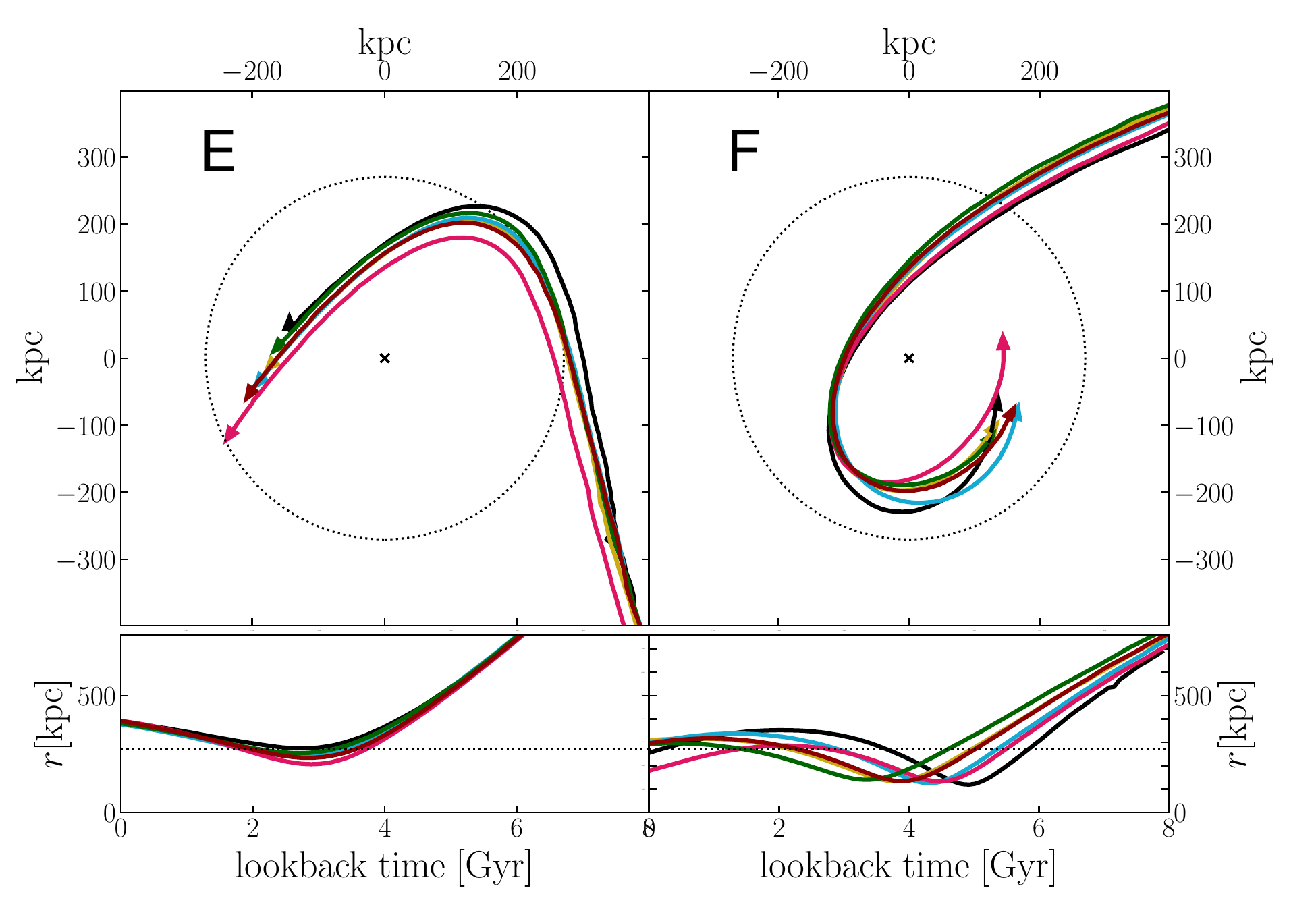}
\end{subfigure}

\caption{Orbital evolution of halos E (left) and F (right). Top panels: projected orbits in the simulation box, centered on the minimum of the host gravitational potential. Bottom panels: orbital radius as a function of time. The dotted lines show the mean of the virial radius of all 6 simulation as an illustration of the extend of the host halo.}
\label{fig:orbits}%
\end{figure}

The orbital evolution of the subhalos is inevitably influenced by the depth of the central gravitational potential of the host galaxy and is therefore indirectly connected to the numerical implementation that governs baryonic mass buildup within the original dark matter–only halo. An illustration is provided in Figure \ref{fig:orbits}.

The upper panels show the projected orbits of halos E (left) and F (right) in the simulation box, centered on the minimum of the host gravitational potential. The lower panels show the corresponding orbital radii as a function of time, as in the top panels of Figure \ref{fig:evo}. Halo F follows a more eccentric, nearly radial orbit with a smaller pericentric distance and lower impact parameter, leading to stronger tidal interactions with the host. In contrast, halo E remains on a more extended, less eccentric orbit with a larger pericentric distance, resulting in weaker tidal forces and reduced stripping, as discussed above.

\end{document}